\documentclass[aps,showpacs,nofootinbib,superscriptaddress]{revtex4}
\usepackage{natbib}
\usepackage{amsbsy,amsmath,amsfonts,bm}
\usepackage{graphicx}
\usepackage{wick}
\usepackage{float}
\usepackage{rotating}

\newcommand{\ignore}[1]{}

\newcommand{\rL}{{\rm L}}

\newcommand{\MeV}{\,{\mathrm{MeV}}}
\newcommand{\GeV}{\,{\mathrm{GeV}}}
\newcommand{\Eq}[1]{{Eq.~(\ref{eq:#1})}}
\newcommand{\SU}{{\mathrm{SU}}}
\newcommand{\vsigma}{{\bm{\sigma}}}
\newcommand{\tr}{{\mathrm{tr}}}

\begin{document}
\title {
Low-lying even parity meson resonances and spin-flavor symmetry revisited
}
\author{C.~Garc\'{\i}a-Recio}
\affiliation{Departamento de F\'{\i}sica
At\'omica, Molecular y Nuclear and Instituto Carlos I de F\'{\i}sica
Te\'orica y Computacional, Universidad de Granada, E-18071
Granada, Spain}
\author{L.~S.~Geng}
\affiliation{School of Physics and 
Nuclear Energy Engineering, Beihang University, Beijing 100191, China}
\author{J.~Nieves}
\affiliation{Instituto de F\'{\i}sica Corpuscular (IFIC), 
Centro Mixto CSIC-Universidad de Valencia,
Institutos de Investigaci\'on de Paterna, Apartado 22085, E-46071
Valencia, Spain}
\author{L.~L.~Salcedo}
\affiliation{Departamento de F\'{\i}sica
At\'omica, Molecular y Nuclear and Instituto Carlos I de F\'{\i}sica
Te\'orica y Computacional, Universidad de Granada, E-18071
Granada, Spain}
\author{En~Wang}
\affiliation{Instituto de F\'{\i}sica Corpuscular (IFIC), 
Centro Mixto CSIC-Universidad de Valencia,
Institutos de Investigaci\'on de Paterna, Apartado 22085, E-46071
Valencia, Spain}
\author{Ju-Jun~Xie}
 \affiliation{Institute of Modern Physics, Chinese Academy of Sciences, Lanzhou 730000, China}
\affiliation{State Key Laboratory of Theoretical Physics, Institute of Theoretical Physics, Chinese Academy of Sciences, Bejing 100190,
China}
\begin{abstract}
We review and extend the model derived in Phys.\ Rev.\ D {\bf 83}
016007 (2011) to address the dynamics of the low-lying even parity
meson resonances. This model is based on a coupled channels
spin-flavor extension of the chiral Weinberg-Tomozawa Lagrangian. This
interaction is then used to study the $S$-wave meson-meson scattering
involving members not only of the $\pi$-octet, but also of the
$\rho$-nonet. In this work, we study in detail the structure of the
SU(6) symmetry breaking contact terms that respect (or softly break)
chiral symmetry. We derive the most general local (without involving 
derivatives) terms consistent with the chiral symmetry breaking
pattern of QCD. After introducing sensible simplifications to reduce
the large number of possible operators, we carry out a
phenomenological discussion of the effects of these terms. We show how
the inclusion of these pieces leads to an improvement of the
description of $J^P = 2^+$ sector, without spoiling the main features
of the predictions obtained in the original model in the $J^P = 0^+$
and $J^P = 1^+$ sectors. In particular, we find a significantly better
description of the $I^G(J^{PC})=0^+(2^{++}), 1^-(2^{++})$ and the
$I(J^{P})=\frac{1}{2}(2^{+})$ sectors, which correspond to the
$f_2(1270)$, $a_2(1320)$ and $K^*_2(1430)$ quantum numbers,
respectively. \ignore{Thus, we end up with a quite robust and
  successful scheme to study the low-lying even parity resonances that
  are dynamically generated by the logs that appear in the unitarity
  loops.}
\end{abstract}
\pacs{
11.10.St Bound and unstable states; Bethe-Salpeter equations,
13.75.Lb Meson-meson interactions,
14.40.Rt Exotic mesons,
14.40.Be Light mesons (S=C=B=0)
}
\date{\today}
\maketitle
\tableofcontents
\section{Introduction}

The study of the lowest--lying hadron resonances dynamics has received
a lot of attention in the last decades, in particular since it was
realized that some of them cannot be easily accommodated as radial or
angular excitations of the Constituent Quark Model (CQM) ground
states. Some examples are the low-lying scalar $f_0(500)$, $f_0(980)$,
$a_0(980)$ and $K^*_0(800)$, or axial vector $a_1(1260)$, $b_1(1235)$,
$h_1(1170)$, $f_1(1285) $, $K_1(1270)$ mesons.  The field has
experimented a considerable boost in the last five years, because
several clear candidates for exotic states can be found among the
recently discovered hidden bottom and charm $XYZ$ resonances reported
by the Belle, BABAR, $D0$ and CDF collaborations\footnote{For
  instance, the isovector $J^{PC}= 1^{+-}$ $Z_b(10610)$ and $Z_b(10
  650)$ resonances (which are located just a few MeV above the $B\bar
  B^*$ and $B^*\bar B^*$ thresholds, respectively~\cite{Belle:2011aa})
  or the $1^{++}$ isoscalar hidden charm state $X(3872)$ placed close
  to the $D^0\bar D^{^0*}$ threshold~\cite{Choi:2003ue}.}.  There has
been a steady activity in the context of CQM's aiming to supplement
these models with exotic components due to existence of tetraquark
degrees of freedom inside of the hadrons (see for instance the
discussion in Ref.~\cite{Vijande:2007ix}).  Such components might lead
to extended CQM schemes where the known exotic resonances could be
generated and their main features be described. Here, however, we will
pay attention to a different approach, in which the hadron resonances
appear as bound or resonant states of an interacting pair of ground
state hadrons (mesons of the $\pi$ octet and the $\rho$ nonet and
baryons of the $N$ octet and $\Delta (1232)$ decuplet, when the study
is limited to the three lightest quark flavors).  In this molecular
picture, hadron resonances show up as poles in the First or Second
Riemann Sheets (FRS/SRS) of certain hadron--hadron amplitudes. The
positions of the poles determine masses and widths of the resonances,
while the residues for the different channels define the corresponding
coupling or branching fractions\footnote{Some studies have also
  adopted an hybrid approach performing coupled channels calculations
  including quark model and molecular configurations (see for instance
  the discussion in Ref.~\cite{Ortega:2009hj}).}. The interaction 
among the ground state hadrons is thus the first ingredient to build
this molecular scheme. These are usually obtained from Effective Field
Theories (EFT's) that incorporate constrains deduced from some
relevant exact or approximate symmetries of Quantum Chromodynamics
(QCD). In this context, it is clear that Chiral Perturbation Theory
(ChPT)~\cite{Weinberg:1978kz,Gasser:1983yg,Gasser:1984gg} and Heavy
Quark Spin Symmetry (HQSS)~\cite{IW89,Ne94,MW00} should play relevant
roles, when designing interactions involving Goldstone bosons or
charm/bottom hadrons, respectively. In this work, we will focus in the
light SU(3) flavor sector and we will leave the extension of this
discussion to heavy molecules for future research.

ChPT is a systematic implementation of chiral symmetry and of its pattern of
spontaneous and explicit breaking, and it provides a model independent scheme
where a large number of low-energy non-perturbative strong-interaction phenomena
can be understood.  It has been successfully applied to study different
processes involving light ($u$ and $d$) or strange ($s$) quarks. Because ChPT
provides the scattering amplitudes as a perturbative series, it cannot
describe non-analytical features as poles. Thus, ChPT cannot directly describe the
nature of hadron resonances.  In recent years, it has been shown that by
unitarizing the ChPT amplitudes in coupled channels, the region of application
of ChPT can be greatly extended. This approach, commonly referred as Unitary
Chiral Perturbation Theory (UChPT), has received much attention and provided
many interesting results, in particular in the meson-meson sector where we
will concentrate our attention in this work,~\cite{Truong:1988zp,
  Dobado:1989qm, Dobado:1992ha, Dobado:1996ps, Hannah:1997ux, Oller:1997ng,
  Oller:1997ti, Nieves:1998hp, Kaiser:1998fi, Oller:1998hw,
  Oller:1998zr,Nieves:1999bx, Nieves:2001de, GomezNicola:2001as, Lutz:2003fm,
  Roca:2005nm, Geng:2006yb, GomezNicola:2007qj, Molina:2008jw, Geng:2008gx,
  Albaladejo:2008qa, GarciaRecio:2010ki}.  It turns out that many meson-meson
resonances and bound states appear naturally within UChPT. These states are
then interpreted as having ``dynamical nature.'' In other words, they are not
genuine $q\bar{q}$ states, but are mainly built out of their meson-meson
components\footnote{The situation is similar in the meson-baryon sector, for
  recent works there see Refs.~\cite{Gamermann:2011mq, Bruns:2010sv}.}.  To
distinguish among these two pictures, it has been suggested to follow the
dependence on a variable number of colors $N_C (> 3)$  of the resonance
properties by assuming that hadronic properties scale similarly as if $N_C$
was large. Some interesting results are being obtained from this
perspective~\cite{Pelaez:2003dy, Pelaez:2006nj, Geng:2008ag, Nieves:2009ez,
  Nieves:2009kh, Nieves:2011gb, Guo:2011pa}, and at present there exists some
controversy on the nature of the $f_0(500)$ resonance~\cite{Pelaez:2006nj,
  Nieves:2011gb, Guo:2011pa} for which accurate models are available.

The present work is an update of Ref.~\cite{GarciaRecio:2010ki}, where
was derived a spin-flavor extension of chiral symmetry to study
the $S$-wave meson-meson interaction involving members not only of the
$\pi$-octet, but also of the $\rho$-nonet. The similar approach for
baryon-meson dynamics was initiated in \cite{GarciaRecio:2005hy,
GarciaRecio:2006wb}.  Elastic unitarity in
coupled channels is restored in \cite{GarciaRecio:2010ki} by solving a
renormalized coupled-channel Bethe--Salpeter Equation (BSE) with an
interaction kernel deduced from spin-flavor extensions of the ChPT
amplitudes. In the scheme of Ref.~\cite{GarciaRecio:2010ki}, the
spin-flavor symmetry was explicitly broken to account for physical
masses and decay constants of the involved mesons, and also when the
amplitudes were renormalized. Nevertheless, the underlying SU(6)
symmetry was still present and served to organize the set of even
parity meson resonances found in that work. Indeed, it was shown that
most of the low-lying even parity PDG (Particle Data Group
Collaboration~\cite{Beringer:1900zz}) meson resonances, specially in
the $J^P=0^+$ and $1^+$ sectors, could be classified according to
multiplets of the SU(6) spin-flavor symmetry group. However, some
resonances, like the isoscalar $f_0(1500)$ or $f_1(1420)$ states,
could not be accommodated within this scheme and it was claimed that
these states could be clear candidates to be glueballs or
hybrids~\cite{GarciaRecio:2010ki}.

Chiral symmetry (CS), and its breaking pattern, is encoded in the
approach of Ref.~\cite{GarciaRecio:2010ki} at leading order (LO) by
means of the Weinberg-Tomozawa (WT) soft pion
theorem~\cite{Weinberg:1966kf,Tomozawa:1966jm}.  This CS input
strongly constraints the pseudoscalar--pseudoscalar ($PP$) and
pseudoscalar--vector ($PV$) channels, since the mesons of the pion
octet were identified with the set of Nambu-Goldstone bosons that
appear for three flavors (due to the spontaneous breaking of
CS). Thus, the main features (masses, widths, branching fractions and
couplings) of the lowest nonet of $S$-wave scalar resonances
($f_0(500)$, $f_0(980)$, $a_0(980)$ and $K^*_0(800)$) found in
Ref.~\cite{GarciaRecio:2010ki} do not significantly differ from those
obtained in previous SU(3) UChPT
approaches~\cite{Oller:1997ti,Oller:1998hw,GomezNicola:2001as}. This
is because these resonances are generated from the interaction of
Nambu-Goldstone bosons, and the influence of the vector--vector ($VV$)
components in these states is small.

The $PV$ and $VV$ sectors have been also systematically studied in
Refs.~\cite{Roca:2005nm} and \cite{Molina:2008jw,Geng:2008gx},
respectively. These works adopt the formalism of the hidden gauge interaction
for vector mesons~\cite{Bando:1984ej,Bando:1987br}.\footnote{Strictly
  speaking, the study of axial-vector resonances carried out in
  Ref.~\cite{Roca:2005nm} does not use the hidden gauge formalism. There, a
  contact WT type Lagrangian is employed. However, the tree level amplitudes
  so obtained coincide with those deduced within the hidden gauge formalism,
  neglecting $q^2/m_V^2$ in the $t$-exchange
  contributions~\cite{Nagahiro:2008cv} and considering only the propagation of
  the time component of the virtual vector mesons.}  In the $PV\to PV$ sector,
as mentioned above, CS constrains the interactions, and the interactions
derived in Ref.~\cite{GarciaRecio:2010ki} and those used in
Ref.~\cite{Roca:2005nm} totally agree at LO in the chiral expansion, despite
their different apparent structure and origin. As a consequence, the results
of Ref.~\cite{GarciaRecio:2010ki} are in general in good agreement with those
previously obtained in Ref.~\cite{Roca:2005nm}, which among others include the
prediction of a two pole structure for the $K_1(1270)$
resonance~\cite{Geng:2006yb}. However, the simultaneous consideration of $PV$
and $VV$ channels made the approach of Ref.~\cite{GarciaRecio:2010ki}
different from that followed in Ref.~\cite{Roca:2005nm} in few cases. One of
the most remarkable cases was that of the $h_1 (1595)$ resonance, which was
dynamically generated for the first time in the work of
Ref.~\cite{GarciaRecio:2010ki}.  The interference $PV \to VV$ amplitudes
turned out to play a crucial role in producing this state in
\cite{GarciaRecio:2010ki}, and that is presumably the reason why the $h_1
(1595)$ resonance was generated neither in the $PV\to PV$ study of
\cite{Roca:2005nm}, nor in the $VV\to VV$ scheme of
Ref.~\cite{Geng:2008gx}. Possibly, the situation is similar for the
$K_1(1650)$ state. These two resonances helped to envisage a clearer SU(6)
pattern in ~\cite{GarciaRecio:2010ki}, which is also followed to some extent
in nature, and that is missed in the separate works of
Refs.~\cite{Roca:2005nm} and ~\cite{Geng:2008gx}.

In general terms, the model of Ref.~\cite{GarciaRecio:2010ki} provides a
fairly good description of the $J^P = 0^+$ and $J^P = 1^+$ sectors. However,
from a phenomenological point of view, the model of
Ref.~\cite{GarciaRecio:2010ki} led to a much poorer description of the $J^P =
2^+$ sector, which for $S$-wave is constructed out of $VV$
interactions. Indeed, the well established $f_2(1270)$ and $K^*_2(1430)$
resonances are difficult to accommodate in the scheme, which needs to be
somehow pushed to its limits of validity.  The hidden gauge interaction for
vector mesons model used in~\cite{Geng:2008gx} seems to be more successful in
describing the properties of the $f_2(1270)$ and $K^*_2(1430)$
resonances. This latter model and that of Ref.~\cite{GarciaRecio:2010ki} are
related for $PV\to PV$ scattering, thanks to CS, but they are completely
unrelated in the $VV$ sector.

The SU(6) spin-flavor symmetry is severely broken in nature. Certainly
it is mandatory to take into account mass breaking effects by using
different pseudoscalar and vector mesons masses. However, this cannot
be done by just using these masses in the kinematics of the amplitudes
derived in a straight SU(6) extension of the WT Lagrangian, since this
would lead to flagrant violations of the soft pion theorems in the
$PV\to PV$ sector due to the large vector meson masses. Instead in
~\cite{GarciaRecio:2010ki}, a proper mass term was added to the
extended WT Lagrangian that produced different pseudoscalar and vector
meson masses, {\em while preserving, or softly breaking, chiral
  symmetry}.  Such term, besides providing masses to the vector
mesons, gives rise to further contact interaction terms
(local). However, some other local interaction SU(6) symmetry breaking
terms respecting (softly breaking) CS can be designed, as we show in
this work. The nature of the contact terms can only be fully unraveled
by requiring consistency with the QCD asymptotic behavior at high
energies~\cite{Ecker:1989yg}, which is far from being trivial. As an
alternative, we will present here a phenomenological analysis of the
effects on the resonance spectrum due to the inclusion of new VV
interactions consistent with CS. Thus, in first place, we will find in
this work the most general four meson-field local (involving no
derivatives) terms consistent with the chiral symmetry breaking
pattern of QCD, and constructed by using a single trace, in the spirit
of the large $N_C$ expansion. Next, we will show that the inclusion of
these pieces leads to a considerable improvement of the description of
$J^P = 2^+$ sector, without spoiling the main features of the
predictions obtained in Ref.~\cite{GarciaRecio:2010ki} for the $J^P =
0^+$ and $J^P = 1^+$ sectors.

The paper is organized as follows. First, we briefly review in
Sect.~\ref{sec:su6old} the model derived in Ref.~\cite{GarciaRecio:2010ki},
including a brief discussion (Subsect.~\ref{sec:bse}) on the BSE in coupled
channels, and the renormalization scheme used to obtain finite amplitudes.
Next in Sect.~\ref{sec:su6new}, we study the interplay between the SU(6)
symmetry breaking local terms and CS, and design two new interaction
terms. Their phenomenological implications are studied in Sect.~\ref{sec:res}.
There, we present results in terms of the unitarized amplitudes and search for
poles on the complex plane. We discuss the results sector by sector trying to
identify the obtained resonances or bound states with their experimental
counterparts~\cite{Beringer:1900zz}, and compare our results with earlier
studies, in particular that of Ref.~\cite{GarciaRecio:2010ki}.  A brief
summary and some conclusions follow in Sect.~\ref{sec:concl}. In Appendix
\ref{app:chiral}, we show that there are just three chiral invariant four meson
contact interactions, if only a single trace is allowed to construct them.  In Appendix
\ref{app:tables} the various potential matrices derived in this work are
compiled for the different hypercharge, isospin and spin sectors.

\section{SU(6) extension of the SU(3)-flavor Weinberg-Tomozawa
  Lagrangian}

\label{sec:su6old}

In this section, we briefly review the model derived in
Ref.~\cite{GarciaRecio:2010ki} to describe the $S$-wave interaction of four
mesons of the $\pi$-octet and/or $\rho$-nonet. 

\subsection{The interaction}

In Ref.~\cite{GarciaRecio:2010ki}, the BSE was solved by using as a kernel the
amplitude ${\mathcal H}$ given by Eq.~(40) of
Ref.~\cite{GarciaRecio:2010ki}. This amplitude consists of three different
contributions. Two of them (${\mathcal D}_{\mathrm{kin}}$ and ${\mathcal
  D}_a$) come from the straight extension to SU(6) of the kinetic part of the
LO WT SU(3)--flavor interaction, while the third one, ${\mathcal D}_m$, is
originated by the mechanism implemented in \cite{GarciaRecio:2010ki} to give
different masses to pseudoscalar and vector mesons. To give mass to the vector
mesons certainly requires breaking SU(6) in the Lagrangian, not only
through mass terms but also by interaction terms, due to chiral symmetry.

The lowest-order chiral Lagrangian describing the interaction of pseudoscalar
Nambu-Goldstone bosons is \cite{Gasser:1984gg}
\begin{equation}
 \mathcal{L}=\frac{f^2}{4}\tr\left(\partial_\mu U^\dagger
 \partial^\mu U+\mathcal{M}(U+U^\dagger-2)\right),
\label{eq:chpt}
\end{equation}
where $f\sim 90\MeV$ is the chiral-limit pion decay constant, $U=e^{{\rm
    i}\sqrt{2}\Phi/f}$ is a unitary $3\times 3$ matrix that transforms under
the linear realization of SU(3)$_\rL\otimes$SU(3)$_{\rm R}$, with
\begin{equation}
 \Phi=\left(\begin{array}{ccc}
             \frac{1}{\sqrt{6}}\eta+\frac{1}{\sqrt{2}}\pi^0 & \pi^+ & K^+\\
              \pi^- & \frac{1}{\sqrt{6}}\eta-\frac{1}{\sqrt{2}}\pi^0& K^0\\
               K^- & \bar{K}^0 & -\sqrt{\frac{2}{3}}\eta 
            \end{array}
\right)
,
\label{eq:su3phi}
\end{equation}
and the mass matrix
$\mathcal{M}=\mathrm{diag}(m_\pi^2,m_\pi^2,2m_K^2-m_\pi^2)$ is determined by
the pion and kaon meson masses.

The straight SU(6) extension of \Eq{chpt} from SU(3) to SU(6)
is~\cite{GarciaRecio:2010ki} 
\begin{equation}
 \mathcal{L}_\mathrm{SU(6)}=
\frac{f^2_6}{4}\tr\!\left(\partial_\mu U^\dagger_6
 \partial^\mu U_6+\mathcal{M}_6(U_6+U_6^\dagger-2)\right), \qquad U_6=e^{{\rm
    i}\sqrt{2}\Phi_6/f_6} 
.
\label{eq:lsu6}
\end{equation}
where $U_6$ is now a unitary $6\times 6$ matrix that transforms under the
linear realization of SU(6)$_\rL\otimes$SU(6)$_{\rm R}$.  The Hermitian matrix
$\Phi_6$ is the meson field in the {\bf 35} irreducible representation of
SU(6), and $f_6= f/\sqrt{2}$~\cite{GarciaRecio:2006wb}.  SU(6) spin-flavor
symmetry allows to assign the vector mesons of the $\rho$ nonet and the
pseudoscalar mesons of the $\pi$ octet in the same ({\bf 35}) SU(6)
multiplet. A suitable choice for the $\Phi_6$ field is\footnote{Matrices,
  $A^i_j$, in the dimension 6 space are constructed as a direct product of
  flavor and spin matrices. Thus, an SU(6) index $i$, should be understood as
  $i \equiv (\alpha,\sigma)$, with $\alpha= 1,2,3$ and $\sigma=1,2$ running
  over the (fundamental) flavor and spin quark degrees of freedom,
  respectively.}
\begin{equation}
\Phi_6 = \underbrace{P_a \frac{\lambda_a}{\sqrt{2}}\otimes
  \frac{I_{2\times 2}}{\sqrt{2}}}_{{\displaystyle {\Phi_P}}} + \underbrace{R_{ak}
\frac{\lambda_a}{\sqrt{2}} \otimes \frac{\sigma_k}{\sqrt{2}} + W_k
\frac{\lambda_0}{\sqrt{2}} \otimes
  \frac{\sigma_k}{\sqrt{2}}}_{\displaystyle{\Phi_V}}, \quad a=1,\ldots,8,\quad
  k=1,2,3 
\end{equation}
with $\lambda_a$ the Gell-Mann and $\vsigma$ the Pauli spin matrices,
respectively, and $\lambda_0= \sqrt{2/3} \,I_{3\times 3}$ ($I_{n\times n}$
denotes the identity matrix in the $n$ dimensional space). $P_a$ are the $\pi,
K, \eta$ fields, while $R_{ak}$ and $W_k$ stand for the $\rho$-vector nonet
fields, considering explicitly the spin degrees of freedom.

The first term\footnote{In what follows, we will refer to it as the kinetic
  term.} in $\mathcal{L}_\mathrm{SU(6)}$ preserves both chiral and spin-flavor
symmetries. The second term breaks explicitly chiral symmetry, and taking for
instance $\mathcal{M}_6=m_6 I_{6\times 6}$, provides a common mass, $m_6$, for
all mesons belonging to the SU(6) {\bf 35} irreducible representation.
However, the SU(6) spin-flavor symmetry is severely broken in nature
and it is indeed necessary to take into account mass breaking effects by using
different pseudoscalar and vector mesons masses.

To this end in Ref.~\cite{GarciaRecio:2010ki}, the following
mass term, which replaces that in \Eq{lsu6}, was considered   
\begin{eqnarray}
\mathcal{L}_\mathrm{SU(6)}^{(m)}
&=& 
\frac{f^2_6}{4} \tr\!\left(
{\mathcal M}(U_6+U_6^\dagger -2)\right)
+\frac{f^2_6}{32} 
\tr\! \left( 
{\mathcal M}^\prime 
(
\vsigma \, U_6 \, \vsigma \, U^\dagger_6
+
\vsigma \, U_6^\dagger \,\vsigma \,
U_6
-6
)
\right)
.
\label{eq:lm}
\end{eqnarray}

Here the matrix ${\mathcal M}$ acts only in flavor space and it is to be
understood as ${\mathcal M}\otimes I_{2\times 2}$, and similarly for
${\mathcal M}^\prime$, so that SU(2)$_{\rm spin}$ invariance is preserved by
these mass matrices. Besides, these matrices should be diagonal in the isospin
basis of \Eq{su3phi} so that charge is conserved. Also, $\vsigma$ stands
for $I_{3\times 3}\otimes \vsigma$.

The first term in $\mathcal{L}_\mathrm{SU(6)}^{(m)}$ is fairly standard. It
preserves spin-flavor symmetry when ${\mathcal M}$ is proportional to the
identity matrix and introduces a soft breaking of chiral symmetry when
${\mathcal M}$ is small. This term gives the same mass to pseudoscalar and
vector mesons multiplets. Note that terms of this type are sufficient to give
different mass to pseudoscalars (e.g. $\pi$ and $K$) when SU($N_F)$ is
embedded into SU($N_F^\prime$) (a larger number of flavors). They are not
sufficient however to provide different $P$ and $V$ masses when SU($N_F$) is
embedded into SU($2N_F$) (spin-flavor).

The second term in $\mathcal{L}_\mathrm{SU(6)}^{(m)}$ only gives mass to the
vector mesons: indeed, if one would retain in $U_6$ only the pseudoscalar
mesons, $U_6$ would cancel with $U_6^\dagger$ (since these matrices would
commute with $\vsigma$) resulting in a cancellation of the whole term. This
implies that this term does not contain contributions of the form $PP$
(pseudoscalar mass terms) nor $PPPP$ (purely pseudoscalar interaction). In
addition, when ${\mathcal M}^\prime$ is proportional to the identity matrix
(i.e., exact flavor symmetry) chiral symmetry is also exactly maintained,
because the chiral rotations of $U_6$ commute with $\vsigma$.  This
guarantees that this term will produce the correct $PV\to PV$ contributions to
ensure the fulfillment of soft pion WT theorem~\cite{Weinberg:1966kf,
  Tomozawa:1966jm} even when the vector mesons masses are not themselves
small.

At order $\Phi_6^2$, the Lagrangian of \Eq{lm} provides proper mass terms for
$P$ and $V$ mesons, while at order $\Phi_6^4$ it gives rise to four meson
interaction terms. In the exploratory study of Ref.~\cite{GarciaRecio:2010ki},
the chiral breaking mass term ($\mathcal{M}$) was neglected, and a common
mass, $m_V$, for all vector mesons ($\mathcal{M}^\prime= m_V^2 I_{3\times
  3}\,I_{2\times 2}$) was used.\footnote{A vector meson nonet averaged mass
  value $m_V= 856\MeV$ was employed in \cite{GarciaRecio:2010ki}.  Note,
  however, that the simplifying choice $\mathcal{M}=0$,
  $\mathcal{M}^\prime=m_V^2$, refers only to the interaction terms derived
  from the Lagrangian $\mathcal{L}_\mathrm{SU(6)}^{(m)}$. For the evaluation
  of the kinematical thresholds of different channels, real physical meson
  masses were used in \cite{GarciaRecio:2010ki}.}  With these simplifications,
the interaction piece deduced from $\mathcal{L}_\mathrm{SU(6)}^{(m)}$
reads~\cite{GarciaRecio:2010ki}
\begin{eqnarray}
\mathcal{L}_\mathrm{SU(6)}^{(m;\, {\rm int} )} &=&
\frac{m_V^2}{8f^2}\tr\!\left( 
\Phi_6^4 
+ \vsigma\,\Phi_6^2 \,\vsigma  \Phi_6^2  
-\frac{4}{3} \vsigma\, \Phi_6 \,\vsigma\, \Phi_6^3 
\right)
.
\label{eq:lmint}
\end{eqnarray}
This term gives rise to the local ${\mathcal D}_m$ contribution to the four
meson amplitude ${\mathcal H}$ in Eq.~(40) of
Ref.~\cite{GarciaRecio:2010ki}. The other two contributions, ${\mathcal
  D}_{\mathrm{kin}}$ and ${\mathcal D}_a$, to ${\mathcal H}$ come from the
first term (kinetic) of $\mathcal{L}_{\SU(6)}$ in \Eq{lsu6}.  In addition, in
Ref.~\cite{GarciaRecio:2010ki} were also considered spin-flavor
symmetry-breaking effects due to the difference between pseudoscalar- and
vector-meson decay constants, and an ideal mixing between the $\omega$ and
$\phi$ mesons (see Subsect. IID of Ref.~\cite{GarciaRecio:2010ki} for some
more details, and Table II for the values of the meson and decay constants
used in the numerical calculations).

\subsection{Scattering Matrix and coupled-channel unitarity} 
\label{sec:bse}
The four meson amplitude ${\mathcal H}$ of Eq.~(40) of
Ref.~\cite{GarciaRecio:2010ki} is used as kernel of the BSE, which is solved
and renormalized for each $YIJ$ (hypercharge, isospin and spin)
sector\footnote{Note that for the $Y=0$ channels, $G$-parity is
  conserved. Thus in the $Y=0$ sectors, the kernel amplitude becomes
  block-diagonal, with each block corresponding to odd and even $G$-parities.} in
the so called {\it on-shell} scheme~\cite{Nieves:1999bx}, $T^{YIJ}$ is given
by
\begin{equation}
T^{YIJ}(s)=\frac{1}{1-V^{YIJ}(s)\,G^{YIJ}(s)}V^{YIJ}(s)
.
\label{eq:t-matrix}
\end{equation}
where $V^{YIJ}(s)$ (a matrix in the coupled-channel space) stands for the
projection of the scattering amplitude, ${\mathcal H}$, in the $YIJ$
sector. The corresponding quantity in the present work is defined below by
Eqs.~(\ref{eq:19}), (\ref{eq:vsu6}) and (\ref{eq:vsu6new}). $\sqrt{s}$ is the center
or mass energy of the initial or final meson pair.  $G^{YIJ}(s)$ is the loop
function and it is diagonal in the coupled-channel space. Suppressing the
indices, it is written for each channel as
\begin{equation}
G(s)=\bar G(s) + G((m_1+m_2)^2)
.
\end{equation}
The finite function $\bar G(s)$ can be found in Eq.~(A9) of
Ref.~\cite{Nieves:2001wt}, and it displays the unitarity right-hand cut of the
amplitude. On the other hand, the constant $G((m_1+m_2)^2)$ contains the
logarithmic divergence. After renormalizing using the dimensional
regularization scheme, one finds
\begin{equation}
G(s=(m_1+m_2)^2) = \frac{1}{16\pi^2} \left ( a(\mu) +
\frac{2}{m_1+m_2} \left \{ m_1 \ln \frac{m_1}{\mu} + m_2 \ln
\frac{m_2}{\mu} \right\}\right)
\label{eq:rs}
\end{equation}
where $\mu$ is the scale of the dimensional regularization. Changes in the
scale are reabsorbed in the subtraction constant $a(\mu)$, so that the results
remain scale independent. Any reasonable value for $\mu$ can be used. In
Ref.~\cite{GarciaRecio:2010ki} $\mu=1\GeV$ was adopted and we take the same
choice in the present work.

Poles, $s_R$, in the SRS of the corresponding BSE scattering amplitudes
($T^{YIJ}(s)$) determine the masses and widths of the dynamically generated
resonances in each $YIJ$ sector (namely $s_R= M^2_R-{\rm i}\ M_R \Gamma_R$).
In some cases, there appear real poles in the FRS of the amplitudes which
correspond to bound states. Finally, the coupling constants of each resonance
to the various meson-meson states ($i,j$ indices) are obtained from the
residues at the pole, by matching the BSE amplitudes to the expression
\begin{equation}
T^{YIJ}_{ij}(s)=\frac{g_i  g_j }{s-s_R} \,,
\label{eq:pole}
\end{equation}
for energy values $\sqrt{s}$ close to the pole. The couplings, $g_i$, are
complex in general.

\section{SU(6) symmetry breaking terms and chiral invariance}

\label{sec:su6new}

Regarding the spin symmetry breaking term of
$\mathcal{L}_\mathrm{SU(6)}^{(m)}$ (the term with ${\mathcal M}'$ in \Eq{lm}),
it should be noted that there is a large ambiguity in choosing it. Being a
contact term, it cannot contain $PPPP$ contributions, due to chiral symmetry,
and for the same reason the terms $PPVV$ are also fixed, as already
noted. However, $VVVV$ terms are not so constrained. One can easily propose
alternative forms for $\mathcal{L}_\mathrm{SU(6)}^{(m)}$ which would still be
acceptable from general requirements but would yield different $VVVV$
interactions. The choice in \Eq{lm} is just the simplest or minimal one.

Let us consider the contact or ultra-local terms, i.e., involving no
derivatives, that can be written down with the desired properties. These
properties include hermiticity, $C$, $P$ and $T$, invariance under rotations
and chiral symmetry. Of course, spin-flavor cannot be maintained, as we want
to give different masses to pseudoscalar and vector mesons.

In the absence of derivatives, the parity transformation is equivalent to $U_6
\to U_6^\dagger$, likewise, $C$ implies $U_6\to U_6^T$ (transposed), and
time-reversal is $U_6\to U_6$ but acting antilinearly. As it turns out, $C$
and $T$ invariances are automatically implied by the other symmetries if the
coupling constants are real (o purely imaginary if $\epsilon_{ijk}$ is
involved).\footnote{Strictly speaking, we cannot invoke the CPT theorem,
  since our interaction is unitary and local but does not have full Lorentz
  invariance. Nevertheless, $T$ turns out to be an automatic consequence of
  $C$ and $P$, and the other assumptions (locality, unitarity and rotational
  invariance).}

Rotational invariance is ensured if the only tensors involved are Pauli
matrices, as well as $\delta_{ij}$ and $\epsilon_{ijk}$.

Under chiral transformations $U_6\to \Omega_L^\dagger U_6 \Omega_R$, where
$\Omega_{L,R}$ are matrices of $\SU(N_F)$, (so actually, they denote
$2N_F\times 2N_F$ matrices of the form $\Omega_{L,R}\times I_{2\times 2}$). Vector
invariance ($\Omega_L = \Omega_R$), the diagonal part of the chiral group, is automatic if the
operators are constructed as traced products of $U_6$, $U_6^\dagger$ and
$\vsigma$ that commutes with the flavor matrices $\Omega_{L,R}$. Note that the
matrix ${\mathcal M}^\prime$ in \Eq{lm} must be a multiple of the identity if
vector invariance is exactly enforced, as we do in this discussion.  Finally,
full chiral invariance requires that $U_6$ and $U_6^\dagger$ blocks should
occupy alternate positions in the trace (cyclically), with Pauli matrices
inserted in between.

A closer look shows that there should be at least one $\vsigma$ between
consecutive $U_6$ and $U_6^\dagger$ (cyclically), and also no more than one
$\vsigma$ is required due to the relation $\sigma_i\sigma_j= \delta_{ij} +
{\mathrm i} \epsilon_{ijk}\sigma_k$. Therefore the total number of $\vsigma$
operators is even and so the number of $\epsilon_{ijk}$ is also even. This
implies that no Levi-Civita tensor $\epsilon_{ijk}$ is needed, due to the
identity
\begin{equation}
\epsilon_{ijk}\epsilon_{abc} = 
\delta_{ia}\delta_{jb}\delta_{kc} + \delta_{ib}\delta_{jc}\delta_{ka}
+ \delta_{ic}\delta_{ja}\delta_{kb} - \delta_{ia}\delta_{jc}\delta_{kb}
- \delta_{ic}\delta_{jb}\delta_{ka} - \delta_{ib}\delta_{ja}\delta_{kc}   
.
\end{equation}
This leads us to the conclusion that the most general contact interaction with
the required symmetries are traced products of blocks
\begin{equation}
\mathcal{U}_{ij} = \sigma_i U_6 \sigma_j U_6^\dagger
,
\end{equation}
that is, products of blocks $\tr(\mathcal{U}_{ij}\mathcal{U}_{kl}\cdots)$,
with the indices contracted in any order.

In principle, there is an infinite number of such interactions
(although relations among them do appear if a concrete number of
flavors, say $N_F=3$, is assumed). Nevertheless, the interaction is
not needed to all orders in the meson field $\Phi_6$, rather only
quadratic and quartic terms need to be retained.\footnote{Parity
  invariance, $\Phi_6\to-\Phi_6$ implies that the interaction contains
  only terms with an even number of meson fields. Of course, this is
  would no longer be true if derivatives were allowed, since this
  would allow anomalous terms involving $\epsilon_{\mu\nu\alpha\beta}$
  \cite{Wess:1971yu,Witten:1983tw}.} Clearly, there is just a finite
number of such $O(\Phi_6^2)+O(\Phi_6^4)$ chiral invariant terms, for
the simple reason that only a finite number of quadratic plus quartic
structures can be written down. Without assuming chiral symmetry there
are $21$ such generic structures (and only $18$ if $N_F=3$ is
specifically assumed). Chiral symmetry imposes relations and reduces
the number from 21 to 10 (9 if $N_F=3$ is assumed). This is a rather
large number of parameters. In order to reduce the problem to a more
manageable size, we will consider here only terms with just a single
trace (rather than products of them). We only mention that such
restriction can be justified from large $N_C$ arguments
\cite{'tHooft:1974hx,Witten:1979kh}. The restriction to a single trace
puts conditions on the possible mass terms for the vector mesons,
specifically $\tr(\Phi_6^2)$ and $\tr((\sigma_i\Phi_6)^2)$ are allowed
but $(\tr(\sigma_i\Phi_6))^2$ is discarded. This implies that the
$\rho$ and $\omega$ mesons cannot be given different masses. Such
degeneracy is very well satisfied experimentally and this gives some
basis to our simplifying assumption.

If only terms with a single trace are retained, the number of possible
quadratic plus quartic operators is 8, and just 3 combinations of them are
chirally invariant. We show this in detail in Appendix \ref{app:chiral}.

The 3 chiral invariant combinations can be obtained by expanding three
independent operators of the type
$\tr(\mathcal{U}_{ij}\mathcal{U}_{kl}\cdots)$ to order $\Phi_4^4$.  Up to two
$\mathcal{U}_{ij}$ blocks and a single trace, only three different operators
can be written down, and they are sufficient for our purposes:
\begin{eqnarray} 
{\mathcal O}_1 &=& \tr\!\left(\mathcal{U}_{ii} - 3 \right), 
\nonumber \\ 
{\mathcal O}_2 &=& \tr\!\left(
\mathcal{U}_{ii} \mathcal{U}_{jj} - 9 \right)
+ \mbox{h.c.},
\label{eq:13}
 \\ 
{\mathcal O}_3 &=& \tr\!\left(
\mathcal{U}_{ij} \mathcal{U}_{ij}  + 3 \right)
.
\nonumber
\end{eqnarray} 
Expanding in the fields, we find
\begin{eqnarray} 
{\mathcal O}_1 &=& 
\frac{3}{f^2} \tr\!\left(
- 4\Phi_6^2 
+ \frac{4}{3} \sigma_i \Phi_6 \sigma_i\Phi_6  
\right)
+ 
\frac{4}{f^4} \tr\!\left( 
\Phi_6^4 
+ \sigma_i\,\Phi_6^2 \,\sigma_i  \Phi_6^2  
-\frac{4}{3} \sigma_i\, \Phi_6 \,\sigma_i\, \Phi_6^3 
\right)
+ {\mathcal O}(\Phi_6^6)
\\
\frac{{\mathcal O}_2-20{\mathcal O}_1}{12} &=& \frac{16}{f^4}
\tr\!\left(
\Phi_6^4 
+ \frac{5}{6}  \sigma_i\,\Phi_6^2 \,\sigma_i \Phi_6^2
-\frac{4}{3} \sigma_i\, \Phi_6 \,\sigma_i \,\Phi_6^3
+ \frac{1}{6}  \Phi_6 \,\sigma_i \,\Phi_6 \,\sigma_i \,\Phi_6 
\,\sigma_j\Phi_6 \,\sigma_j 
\right.
\nonumber \\ &&
+ \left.
\frac{1}{6} \mathrm{i}\epsilon_{ijk} \Phi_6^2 \,\sigma_i \,\Phi_6 \,\sigma_j 
\,\Phi_6 \,\sigma_k
\right) + {\mathcal O}(\Phi_6^6)
,
\\ 
\frac{{\mathcal O}_3}{3} &=& 
\frac{16}{f^4}
\tr\!\left(
\Phi_6^4 -\frac{4}{3} \sigma_i\, \Phi_6 \,\sigma_i \,\Phi_6^3  
+ \frac{1}{3}  \Phi_6 \,\sigma_i \,\Phi_6 \,\sigma_j \,\Phi_6
\,\sigma_i\Phi_6 \,\sigma_j 
- \frac23 \mathrm{i}\epsilon_{ijk} \Phi_6^2 \,\sigma_i \,\Phi_6 
\,\sigma_j \,\Phi_6 \,\sigma_k
\right)
+ {\mathcal O}(\Phi_6^6)
.
\end{eqnarray} 
These three operators are linearly independent. Moreover, we show in
the Appendix
\ref{app:chiral} that, up to order ${\mathcal O}(\Phi_6^4)$, any other
operator arising from the set of chiral invariant Lagrangians
$\tr(\mathcal{U}_{ij}\mathcal{U}_{kl}\cdots)$ can be expressed as a linear
combination of ${\mathcal O}_{1,2,3}$. This is one of the most important
results of this work.

The coupling of the operator ${\mathcal O}_1$ has to be $f^2 m_V^2/32$, to
generate a proper mass term for the vector mesons. This implies
\begin{equation}
{\mathcal L}_1= \frac{f^2 m_V^2}{32} {\mathcal O}_1 = -\frac{1}{2} m_V^2
\tr\!\left(\displaystyle{\Phi^2_V}\right) +
\mathcal{L}_\mathrm{SU(6)}^{(m;\, {\rm int} )}+ {\mathcal O}(\Phi_6^6)
\end{equation}
with $\mathcal{L}_\mathrm{SU(6)}^{(m;\, {\rm int} )}$ given in
\Eq{lmint}. However, a priori we cannot fix the couplings $g_2$ and $g_3$
of the operators ${\mathcal O}_2$ and ${\mathcal O}_3$, which were set
arbitrarily to zero in Ref.~\cite{GarciaRecio:2010ki}. Here, we aim to explore
the physical consequences of keeping these two interaction terms finite. Thus,
we will consider here an additional contact four meson interaction Lagrangian
\begin{equation}
\delta\mathcal{L}_\mathrm{SU(6)}^{(m;\, {\rm int} )} = \frac{f^2
  m_V^2}{64} \left (  \frac{g_2}{4\pi} \frac{{\mathcal
    O}_2-20{\mathcal O}_1}{12} +  \frac{g_3}{4\pi} 
\frac{{\mathcal O}_3}{3}  \right )
\label{eq:deltaL}
\end{equation} 
When the terms above are taken into account, the final $S$-wave four meson
amplitude, ${\mathcal H}$ reads

\begin{eqnarray}
{\mathcal H} &=& {\mathcal H_0}  + \delta{\mathcal H}
\label{eq:19}\\
{\mathcal H_0}&=& \frac{1}{6f^2} \left(3s-\sum_{i=1}^4
q_i^2\right) {\mathcal D}_{\rm kin} +
\frac{m^2_V}{8f^2} {\mathcal D}_m+  
\frac{1}{2f^2} \frac{m_V^4}{ s} {\mathcal D}_a
\label{eq:vsu6}
\\
\delta{\mathcal H}&=& \frac{m^2_V}{16\pi f^2}\left ( g_2{\mathcal D}_2 +
 g_3{\mathcal D}_3 \right) 
\label{eq:vsu6new}
\end{eqnarray}
where $s=(q_1+q_2)^2$ is the usual Mandesltam variable, with $q_1$ and $q_2$
($q_3$ and $q_4$) the four-momenta of the initial (final) mesons.  The first
term, ${\mathcal H_0}$, coincides with the four meson amplitude used in
Ref.~\cite{GarciaRecio:2010ki} (see Eq.~(40) of this reference) as kernel of
the BSE.\footnote{The matrices ${\mathcal D}_{\rm kin}$, ${\mathcal D}_{\rm
    a}$ and ${\mathcal D}_{\rm m}$ are compiled in the Appendix A of
  \cite{GarciaRecio:2010ki}.}  The second term, $\delta{\mathcal H}$, is the
new dynamical input. The physical consequences of this term will be studied in
this work. The coupled-channel matrices ${\mathcal D}_{2,3}$ are obtained from
the Lagrangian $\delta\mathcal{L}_\mathrm{SU(6)}^{(m;\, {\rm int} )}$ in
\Eq{deltaL}, with the convention $-{\mathrm i}\, \delta{\mathcal H} = {\mathrm
  i} \delta\mathcal{L}_\mathrm{SU(6)}^{(m;\, {\rm int} )}$.  These matrices
are compiled in Appendix \ref{app:tables}. We have verified that the sets of
matrices ${\mathcal D}_m$, ${\mathcal D}_2$ and ${\mathcal D}_3$ (computed for
$N_F=3$) are globally linearly independent.  By inspection of these matrices,
it can be checked that in the $J=1$ sector, the new contribution
$\delta{\mathcal H}$ vanishes provided $7g_2+12g_3=0$.

\section{Results and Discussion}
\label{sec:res}
In this section, we will address the
consequences of adding the amplitude $\delta{\mathcal H}$ to the kernel of the
BSE, from a phenomenological point of view. To that end in each $YIGJ$ sector, we will compare the spectrum of
resonances obtained from the pole structure of the renormalized BSE $T$-matrix
in the FRS and SRS with the main properties of the meson states listed in the
PDG~\cite{Beringer:1900zz}. To better isolate the effects of $\delta{\mathcal
  H}$, we will frequently refer also to the previous results obtained in
Ref.~\cite{GarciaRecio:2010ki}.

The main obstacle to carry out the above program is the enormous freedom that
a priori exists for fixing the subtraction constants. This is not only true
for the present model, all schemes that restore unitarity suffer from the same
problem~\cite{Truong:1988zp, Dobado:1989qm, Dobado:1992ha, Dobado:1996ps,
  Hannah:1997ux, Oller:1997ng, Oller:1997ti, Nieves:1998hp, Kaiser:1998fi,
  Oller:1998hw, Oller:1998zr, Nieves:1999bx, Nieves:2001de,
  GomezNicola:2001as, Lutz:2003fm, Roca:2005nm, Geng:2006yb,
  GomezNicola:2007qj, Molina:2008jw, Geng:2008gx, Albaladejo:2008qa,
  GarciaRecio:2010ki}. The origin of this freedom should be traced back to the
renormalization procedure needed to render finite the unitarized amplitudes,
that always involve a non-perturbative re-summation. Since all meson-meson
theories are effective, their renormalization inescapably requires the
introduction of new and undetermined low energy constants (rLECs). For the
interaction of Goldstone bosons, these rLECs can be related to the low energy
constants that appear in the higher order Lagrangian terms of the systematic
chiral expansion~\cite{Truong:1988zp, Dobado:1989qm, Dobado:1992ha,
  Dobado:1996ps, Hannah:1997ux, Oller:1997ng, Nieves:1998hp, Oller:1998hw,
  Nieves:1999bx, Nieves:2001de, GomezNicola:2001as, Lutz:2003fm}, and in some
cases, they might be constrained with other physical observables. However, no
such systematic expansion exists when the involved bosons are vectors, and
consequently, their related rLECs remain to a large extent unconstrained.
Often, the unknown rLECs are tuned to best reproduce the physical properties
of the resonances generated by the non-perturbative unitary re-summation.

In the renormalization scheme followed in \cite{GarciaRecio:2010ki}, the rLECs
are encoded by the subtraction constants $a(\mu)$ that appear in the
expression of the renormalized loop function in \Eq{rs}.  There is
one such constant for each $YIGJ$ sector and for each channel of the
associated coupled channels space. The $a(\mu)$ are free parameters prior to
supplementing more detailed information from QCD. As said, the situation is
similar in the rest of the approaches applied to the study of vector mesons.
A practical solution to the impasse is found in the
literature~\cite{Roca:2005nm, Geng:2006yb, GomezNicola:2007qj, Molina:2008jw,
  Geng:2008gx, GarciaRecio:2010ki}, namely, for $\mu=1\GeV$, the various
$a(\mu)$ are fixed to values around  $-2$. The $a(\mu)$'s are let to vary
around the value $-2$ to best describe the known phenomenology in each $YIGJ$
sector. This default value of $-2$ is suggested from analysis where an
ultraviolet (UV) hard cutoff $\Lambda$ is used to renormalize the loop
function, instead of dimensional regularization. The relation between
the subtraction constant $a(\mu)$ at the scale $\mu$ and $\Lambda$ is
\begin{equation}
a(\mu)= - \frac{2}{m_1+m_2} \left \{ m_1\, \ln \left[
\frac{\Lambda+\sqrt{\Lambda^2 +m_1^2}}{\mu}\right] + m_2 \,\ln \left[
\frac{\Lambda+\sqrt{\Lambda^2 +m_2^2}}{\mu} \right]\right\}
.
\end{equation}
For $\mu\sim 1\GeV$, and assuming a cutoff of the same order of magnitude,
$-2$ turns out to be a natural choice for the subtraction constant
$a(\mu)$.

The idea behind the above choice for the range of variation of the rLECs is to
focus on the resonances whose dynamics is mostly determined by the unitarity
loops. A clear example of one such resonance is the $f_0(500)$, that is
dynamically generated from $\pi\pi$ re-scattering with a cutoff value of the
order of $700\MeV$~\cite{Oller:1997ti}. This translates to $a(\mu=1\GeV)\sim
-0.7$. However, to similarly describe the $\rho$-meson, purely from $\pi\pi$
re-scattering, requires $a(\mu=1\GeV)\sim
-12$~\cite{Nieves:1998hp,Nieves:1999bx}. This would lead to unnatural values
for the UV cutoff, of the order of $200\GeV$ (note the logarithmic
dependence). Actually, the $\rho$-meson is rather insensitive to the chiral
loops and its dynamics is mostly determined by the low energy constants that
appear in the ${\cal O}(p^4)$ chiral Lagrangian~\cite{Nieves:2009ez}.

In our scheme, the mesons of the $\rho$-nonet, used to build the coupled
channels space, are {\it preexisting states} (to adopt the terminology
of~\cite{Oller:1998zr}), rather than dynamically generated from the
re-scattering of Goldstone bosons. In this view, it looks appropriated to
restrict the rLECs to values that could be related to reasonable values of the
UV cutoff. Specifically, $\Lambda$ will be allowed to lie in the interval
$[0.5, 5]\GeV$. Even after this constraint, there is still a large freedom in
varying all different rLECs.

Another ambiguity in the model needs to be fixed, namely, the values of the
two new couplings $g_2$ and $g_3$ in \Eq{vsu6new}, which are totally
undetermined yet. To be practical, we introduce here a further simplification
by imposing the relation
\begin{equation}
g_2=-\frac{12}{7} g_3
.
\end{equation} 
This relation between $g_2$ and $g_3$ guarantees the $J=1$ sectors are not
affected by the new amplitude $\delta{\cal H}$. Note that in the $J=1$ sector,
the $PV\to PV$ terms are constrained by chiral symmetry and that in the
previous analysis of Ref.~\cite{GarciaRecio:2010ki}, where $\delta{\cal H}$
was neglected, this sector was quite successfully described. Among other, the
$\left[I^G(J^{PC})\right] 0^- (1^{+-})$ $h_1(1170)$, $h_1(1380)$ and
$h_1(1595)$, the $1^+ (1^{+-})$ $b_1(1235)$, the $1^- (1^{+-})$ $a_1(1260)$
and $a_1(1640)$ resonances were dynamically generated. Furthermore, the double
pole structure of the $I(J^P)= \frac12 (1^+)$ $K_1(1270)$ resonance, firstly
uncovered in \cite{Roca:2005nm, Geng:2006yb}, was strongly confirmed in
Ref.~\cite{GarciaRecio:2010ki}, as well.

Assuming natural values for $g_3$, we have let this coupling vary in the
interval $[-25,25]$.  For each value of $g_3$, we have looked at the
different $YIG$ sectors for $J=0$ and $J=2$, and have examined the pattern of
generated resonances, when the rLECs $a(\mu)$'s are left to vary in each
particle--channel in the numerical range associated to UV cutoffs comprised in
the interval $[0.5,5]\GeV$. We find that, in general, the $J=0$ sectors are
not much affected by the new couplings. Simultaneously,
values of $|g_3| \le 0.25$  yield a definitely better description of the
main features of the PDG $J^{P}=2^+$ low-lying resonances than that achieved
in Ref.~\cite{GarciaRecio:2010ki}. Specifically, in the results to be
presented below we have taken
\begin{equation}
\frac{g_3 m^2_V}{16\pi f^2} = 0.1
\end{equation}
For this value of $g_3$, a quite good overall description of the different
$J=0,2$ sectors is obtained. Small variations around this value,  keeping $|g_3| \le 0.25$, can be
compensated by the rLECs leading to descriptions of similar quality.

The value of the new coupling is relatively small.  A tentative argument can
be advanced to explain why such small value was to be expected.  In the heavy
quark limit, QCD shows an approximate spin-symmetry~\cite{IW89,Ne94,MW00} that
requires the spin symmetry breaking terms to be suppressed by at least one
power of the heavy quark mass.  Since the operators to which $g_3$ couples
ought to be suppressed in a hypothetical large $m_V$ limit, the natural
combination $(g_3 m^2_V/(16\pi f^2)$ that appears in the Lagrangian should be
of order $\Lambda_{\rm QCD}/m_V$, or $g_3 \sim {\cal O}(\left(\Lambda_{\rm
  QCD}/m_V)^3 \right)$, with $\Lambda_{\rm QCD}\sim 250\MeV$ standing for some
energy scale relevant in the problem, in addition to the averaged vector mass.

In what follows, we will present and discuss the results that we have found in
the various $YIGJ$ sectors.

\subsection{Hypercharge 0, isospin 0 and spin 0}

\begin{table}[htpb]
\caption{Pole positions (in MeV) and moduli of the couplings $|g|$ (in GeV) in the
  $(Y,I,J)=(0,0,0)$ sector that corresponds to the $I^G(J^{PC})=0^+(0^{++})$
  quantum numbers.  The subtraction constant has been set to the default
  values $a=-2.0$ in all channels.  We also compile the results obtained in
  Ref.~\cite{GarciaRecio:2010ki}, where $\delta{\cal H}$ was set to zero, and
  the available information in the PDG on masses and widths (in MeV) of the
  possible counterparts. The channels open for decay have been highlighted in
  boldface.}
\label{tab:000}  \vspace{0.5cm}
\begin{tabular}{c|cccccccc|c|cc}
\hline\hline
  $(M_R,\Gamma_R)$& $\pi\pi$ &  $\bar{K}K$ & $\eta\eta$ & $\rho\rho$ & $\omega\omega$ & $\omega\phi$
  & $\bar{K}^* K^*$ &  $\phi\phi$   & $(M_R,\Gamma_R)$ ~\cite{GarciaRecio:2010ki} & 
\multicolumn{2}{c}{PDG $(M_R,\Gamma_R)$~\cite{Beringer:1900zz}}\\ \hline
$(631,406)$ & {\bf 3.54} & 0.38 & 0.38 & 8.17 & 8.29 & 0.11 & 7.72 & 1.94   &
$(602,426)$ &$f_0(500)$ & $(400\sim 550,400\sim 700)$\\

$(971,0) $ & {\bf 0.03} & 2.49 & 2.06 & 2.96 & 2.12 & 1.42 & 3.34 & 3.01   &
$(969,0) $ & $f_0(980)$ & $(990\pm 20,40\sim 100)$\\

$(1365,124)$ & {\bf 0.53} & {\bf 3.26} & {\bf 0.82} & 1.27 & 2.74 & 7.12 & 10.54 & 10.33
&$(1349,124)$ & $f_0(1370)$& $(1200\sim1500,200\sim500)$\\
$(1729,104)$ & {\bf 0.04} & {\bf 0.83} & {\bf 3.16} & {\bf 0.32} & {\bf 0.39} & 3.27 & 2.38 & 13.66  &
$(1722,104)$& $f_0(1710)$& $(1720\pm 6,135\pm 8)$\\
\hline\hline
\end{tabular}

\end{table}
In this sector we find four poles. Their positions and couplings are compiled
in Table \ref{tab:000}, where we have also collected the pole positions found
in our previous work of Ref.~\cite{GarciaRecio:2010ki}.  Masses and widths
listed in the PDG \cite{Beringer:1900zz} of the possible resonances that could
be identified with these states are also given in the table. As anticipated,
the inclusion of $\delta{\cal H}$ in the present work has very little effect
and the present results are qualitatively similar to those already obtained in
\cite{GarciaRecio:2010ki}. We refer to that work for further details and
grounds on the identification proposed in Table \ref{tab:000}. Very briefly,
the lowest two poles can be easily identified with the $f_0(600)$ and
$f_0(980)$ resonances. There are some differences with other
works~\cite{Oller:1998hw,GomezNicola:2001as} mainly because we have neglected
the pseudoscalar meson mass terms and have included vector meson-vector meson
channels. The identification of the other two poles is less direct, though it
is quite reasonable to associate them to the $f_0(1370)$, and $f_0(1710)$
resonances, as it is argumented in \cite{GarciaRecio:2010ki}. On the other
hand, the $I^G=0^+(J^{PC}=0^{++})$ $f_0(1500)$ resonance cannot be
accommodated within this scheme and thus it would be a clear candidate to be a
glueball or a hybrid.

A final remark concerns the $f_0(500)$, in the most recent update of the
PDG~\cite{Beringer:1900zz}, the traditionally large range of values for the
$f_0(500)$ mass (previously called $f_0(600)$ or $\sigma$) has been
considerably shrunk, thanks to the consideration of recent determinations of
the position of this pole obtained in dispersive
approaches~\cite{Caprini:2005zr, GarciaMartin:2011jx}. As can be seen in Table
\ref{tab:000}, now the mass of the $\sigma$ lies in the interval
$400-550\MeV$. The scheme presented here easily accommodates masses for the
$\sigma$ of the order of $500\MeV$, just by slightly modifying the subtraction
constants (rLECs). Note that for a better comparison with our previous work of
Ref.~\cite{GarciaRecio:2010ki}, in Table \ref{tab:000}, all subtraction
constants have been set to the default values $a=-2.0$, as in this latter
reference. However, using instead $-2.3$ for the $PP$ channels and $-1.1$ for
the $VV$ ones, we find that $M_R$ is about $500\MeV$
and $992\MeV$ for the $f_0(500)$ and $f_0(980)$ poles, respectively, in much
better agreement with the masses listed for these resonances in the last
edition of the PDG. The positions of the other two poles placed at higher
energies are not significantly changed, thus our qualitative discussion of
this sector remains unchanged.

\subsection{Hypercharge 0, isospin 0 and spin 2}
\begin{table}[htpb]
\caption{Same as in Table \ref{tab:000}, but for 
the $(Y,I,J)=(0,0,2)$ sector that corresponds to
the $I^G(J^{PC})=0^+(2^{++})$ quantum numbers. The
subtraction constant has been set to $a=-3.88$ in all channels. } 
\vspace{0.5cm}
\begin{tabular}{c|ccccc|c|cc}
\hline\hline $(M_R,\Gamma_R)$& $\rho\rho$ & $\omega\omega$& $\omega\phi$
&$\bar{K}^* K^*$ &$\phi\phi$ & $(M_R,\Gamma_R)$~\cite{GarciaRecio:2010ki} & \multicolumn{2}{c}{PDG $(M_R,\Gamma_R)$~\cite{Beringer:1900zz}} \\
\hline
$( 1279, 0)$ & 4.40 & 4.08 & 1.74 & 4.08 & 10.20  & $( 1289, 0)$ &
$f_2(1270)$ & $(1275.1\pm1.2,185.1_\mathrm{-2.4}^{+2.9})$ \\
 $(1658,74)$ & {\bf 3.56} & {\bf 1.74} &  2.24 & 5.58 & 4.21 &  $(1783,38)$ &
$f_2(1640)$ & $(1639\pm6,99_\mathrm{-40}^{+60})$ \\
\hline\hline
\end{tabular}
\label{tab:002}
\end{table}

\begin{figure}[tbh]
\centerline{\hspace{-2cm}\includegraphics[height=3.cm]{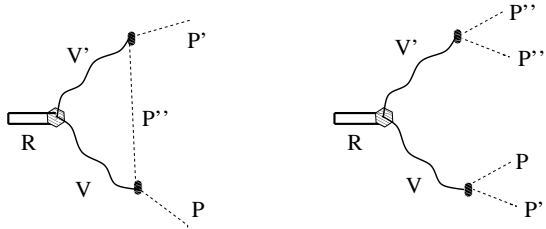}}
\caption{ Resonance ($R$) decay to two (left) or four (right) pseudoscalar
  mesons ($P$, $P^\prime$, $P^{\prime\prime}$, $P^{\prime\prime\prime}$)
  through its $s$-wave (filled pattern hexagon) coupling to two vector mesons
  ($V$,$V^\prime$) and the $p$-wave coupling (black ovals) of these
  latter mesons to two pseudoscalar mesons.}
\label{fig:box}
\end{figure}

In this sector, we find two poles in the SRS/FRS of our amplitudes (see
Table~\ref{tab:002}). We fine-tune a common value of the rLECs to obtain a
mass for the first state (bound) in the vicinity of that quoted in the PDG for
the $f_2(1270)$. Having fixed the rLECs, we find a second pole located at
$(1658,74)\MeV$, with mass and width close to those of the $f_2(1640)$
resonance. Moreover, this second pole has large couplings to the $\rho\rho$,
$\bar{K}^* K^*$ and $\omega\omega$ channels, which will naturally account for
the seen decay modes of the $f_2(1640)$ resonance into $\omega\omega$ and also
into $\bar{K} K$ and $\pi\pi\pi\pi$ through loop mechanisms, like those
depicted in Fig.~\ref{fig:box}.

These loops mechanisms might also provide a sizable width to the first
pole that we have identified with the $f_2(1270)$ resonance.  Indeed,
this resonance is quite broad ($\Gamma\sim 185\MeV$) while in our
case, it appears as a bound state (pole in the FRS) of zero
width. Besides, there exist other mechanisms like $d$-wave $\pi\pi$
decays, which could also be important in this case because the large
available phase space. Note that the $\bar K K$ decay mode ($\sim
$4.5\%) quoted in the PDG for the $f_2(1270)$ resonance can be easily
accommodated in our scheme thanks to the large couplings of our state
to the $\bar K^* K^*$ and $\phi \phi$ channels.

In the hidden gauge model of Ref.~\cite{Geng:2008gx} two states were also
generated in this sector, whose masses agree remarkably well with
those of the
$f_2(1270)$ and $f_2^\prime(1525)$ resonances.\footnote{Thanks to a suitable fine-tuning
  of the subtraction constants.} There, these two resonances appear mostly as
$\rho\rho$ and $\bar K^* K^*$ bound states, respectively. In our case these
channels are still dominant but with a substantial contribution from the
sub-dominant channels. (An exception comes from $\phi\phi$, with a sizable
coupling but a relatively high threshold.) The presence of relatively
important subdominant channels prevents us from identifying the second pole
obtained in our approach with the $f_2^\prime(1525)$ resonance. This is
because this resonance has the distinctive feature of having a very small
branching fraction into the $\pi\pi$ channel ($\sim 0.8\%$) what seems hard to
accommodate with the sizable $\rho\rho$ coupling of our state. On the other
hand, the $\bar K K$ mode, that we expect to be dominant for our second state,
has not been seen in the decays of $f_2(1565)$ and $f_2(1810)$
resonances. This finally brings us to identify our second pole in
Table~\ref{tab:002} with the $f_2(1640)$ resonance.

A final remark is in order here. In our previous analysis carried out in
Ref.~\cite{GarciaRecio:2010ki}, we also found two poles and made the same
identifications as here. However in that work, we could not fine tune the
subtraction constants to obtain masses for the second state below
$1.75\GeV$. Thanks to the consideration of $\delta{\cal H}$ in the present
analysis we have been able to predict two resonances with the appropriate
masses to be identified with the $f_2(1270)$ and $f_2(1640)$
states. Nevertheless to achieve this, we have needed to use values of the
subtraction constant of around $-3.9$. This in turn implies large UV cutoff
values of around $3.5\GeV$, somehow in the limit of what one would expect for
resonances dynamically generated by the unitarity loops driven by the LO
potential. These large UV cutoffs might signal some resemblance between the
nature of these states and that of the so called {\it preexisting states},
like the $\rho$ meson, for which higher order corrections (not driven by
unitarity) in the potential play an important role in their
dynamics~\cite{Nieves:2009ez}.

\subsection{Hypercharge 0, isospin 1 and spin 0}
\begin{table}[htpb]
\caption{Same as in Table \ref{tab:000}, but for 
the $(Y,I,J)=(0,1,0)$ sector that corresponds to the
$I^G(J^{PC})=1^-(0^{++})$ quantum numbers. The subtraction constants have
been fixed to $a=-3.5$ in the $PP$ channels and to $a=-2.16$ in the
$VV$ channels. }
\label{tab:010}\vspace{0.5cm}
\begin{tabular}{c|ccccc|c|cc}
\hline\hline
 $(M_R,\Gamma_R)$ & $\eta\pi$ & $\bar{K}K$ & $\omega\rho$ & $\phi\rho$&
$\bar{K}^*K^*$ & $(M_R,\Gamma_R)$~\cite{GarciaRecio:2010ki}
  & \multicolumn{2}{c}{PDG $(M_R,\Gamma_R)$~\cite{Beringer:1900zz}} \\
  \hline
 $(970, 90)$ & {\bf 2.67} & 2.82 & 7.33 & 2.62 & 0.09 &$(990, 92)$ &
  $a_0(980)$ & $(980\pm20,50\sim100)$    \\
 $(1493,86)$ & {\bf 1.68} & {\bf 2.26} & 2.04 & 8.24 & 8.08 & $(1442,10)$ &
  $a_0(1450)$ & $(1474\pm19,265\pm13)$ \\
 &&&&& & $(1760,24)$ &  \\
\hline\hline
\end{tabular}

\end{table}

There are five coupled channels in this sector: $\pi\eta$, $\bar{K} K$,
$\rho\omega$, $\rho\phi$ and $\bar{K}^*K^*$ and we now find two poles in the
SRS of the amplitudes. Our results are compiled in Table~\ref{tab:010}.  The
lowest pole should be identified with the $a_0(980)$, which has been obtained
in all previous studies considering only pseudoscalar-pseudoscalar coupled
channels. In our scheme, its couplings to the $\pi\eta$ and $K\bar{K}$ are
large, in agreement with the results of earlier studies and with the data, but
it also presents large couplings to the heaviest vector channels, $\omega\rho$
and $\phi\rho$. When comparing to the results of
Ref.~\cite{GarciaRecio:2010ki}, we see that the couplings of this resonance to
vector channels have drastically changed, though these heavier channels have
little influence on the position of this lowest pole and on its allowed decay
modes.

The pole at $(M_R,\Gamma_R)=(1493,86)\MeV$ can be naturally associated to the
$a_0(1450)$ and its main features are similar to those found in
\cite{GarciaRecio:2010ki}.  It decays to $\pi\eta$ and $K\bar{K}$, which is in
agreement with the data. Its large couplings to the vector channels, whose
thresholds are now closer, will give rise to new significant $\omega \pi\pi$ and
$\bar K K \pi\pi$ decay modes, and to an important enhancement of its width
thanks to the broad spectral functions of the $\rho$ and $K^*$ resonances.

Finally, as can be seen in the table, in our previous analysis of
Ref.~\cite{GarciaRecio:2010ki}, we found a third pole, located at
$(M_R,\Gamma_R)=(1760, 24)\MeV$, and whose dynamics was mostly dominated by the
vector channels. This further state could not be associated to any known
state, since the PDG only reports two $a_0$ resonances below $2\GeV$.
Nevertheless, in Ref.~\cite{GarciaRecio:2010ki} we suggested with some
cautions that this third pole, though placed quite below, might be identified
with the very broad $a_0(2020)$ resonance ($\Gamma = 330 \pm 75\MeV$). This
latter resonance is not firmly established at all, and needs further
confirmation. (Indeed, it appears in the section of {\it Further states} of
the PDG.) In the present re-analysis, where the new spin symmetry breaking
terms contained in $\delta {\cal H}$ have been included, this state is no
longer dynamically generated.

\subsection{Hypercharge 0, isospin 1 and spin 2}

\begin{table}[htpb]
\caption{Same as in Table \ref{tab:000}, but for 
the $(Y,I,J)=(0,1,2)$ sector that corresponds to the
$I^G(J^{PC})=1^-(2^{++})$ quantum numbers. The
subtraction constant has been set to $a=-3.88$ in all channels.} \label{tab:012}

\vspace{0.5cm}
\begin{tabular}{c|ccc|c|cc}
\hline\hline
 $(M_R,\Gamma_R)$ & $\omega\rho$ & $\phi\rho$  & $\bar{K}^* K^*$ &
$(M_R,\Gamma_R)$~\cite{GarciaRecio:2010ki} & \multicolumn{2}{c}{PDG
  $(M_R,\Gamma_R)$~\cite{Beringer:1900zz}}  \\
 \hline
$(1319,0)$ & 8.41 & 1.92 & 5.32 & $(1228,0)$ & $a_2(1320)$& $(1318.3_\mathrm{-0.6}^{+0.5},107\pm5)$ \\
$(1747,12)$ & {\bf 1.55} & 3.48 & 4.82 & $(1775,12)$ & $a_2(1700)$& $(1732\pm16,194\pm40)$  \\
\hline\hline
\end{tabular}
\end{table}

There are three coupled channels in this sector: $\bar{K}^* K^*$,
$\omega\rho$, and $\phi \rho$, and we find two poles, one in the FRS and the
other one in the SRS of the amplitudes (see Table~\ref{tab:012}), which might
be associated to the $a_2(1320)$ and $a_2(1700)$ resonances. The first state,
bound in our model, couples strongly to the $\omega\rho$ channel, and its
couplings would give rise to the observed $\pi\pi\pi$ and $\omega\pi\pi$ decay
modes of the $a_2(1320)$ thanks to the width of the virtual $\rho$ meson. In
our previous analysis~\cite{GarciaRecio:2010ki}, we could not place the mass
of this state above $1230\MeV$, despite we tried some fine-tuning of the
subtraction constants.  Thus here, as it was also the case in the
$(Y,I,J)=(0,0,2)$ sector, we find a better overall description of the sector
thanks to the inclusion of the $\delta{\cal H}$ terms.

The $a_2(1320)$ resonance is not dynamically generated in the hidden gauge
model of Ref.~\cite{Geng:2008gx}, though there it is reported one state whose
features are similar to those of the second (heaviest) pole found here.  This
second pole might be associated to the $a_2(1700)$, since its mass and
expected decays into $\omega\rho$, $\omega \pi^-\pi^0$ and $K\bar{K}$ (from
the decays into virtual $\phi\rho$ or $K^*\bar{K}^*$ pairs) are in good
agreement with the information listed in the PDG for the $a_2(1700)$
resonance. However, the state predicted here, as it was the case in
Refs.~\cite{GarciaRecio:2010ki,Geng:2008gx} turns out to be much narrower than
this resonance. This is probably an indication that other mechanisms, such as
coupled-channel $d$-wave dynamics, might play an important role in this
case. Nevertheless, there exists a large uncertainty in the experimental
status of the $a_2(1700)$.

Finally, we should point out here that in this sector, we have also needed to
make use of large UV cutoffs ($\sim 3.5\GeV$), somehow in the limit of what
one would expect for resonances dynamically generated by the unitarity loops
driven by the LO potential.

\subsection{Hypercharge 1, isospin 1/2 and spin 0}

In this sector we find three poles, with positions and couplings compiled in
Table \ref{tab:1-12-0}. There we have also collected the pole positions found
in our previous work of Ref.~\cite{GarciaRecio:2010ki}.  Masses and widths
listed in the PDG \cite{Beringer:1900zz} of the possible resonances that could
be identified with these states are also given in the table. As in the
$(Y,I,J)=(0,0,0)$ sector, the inclusion of $\delta{\cal H}$ in the present
work has very little effect, and the present results are quite similar to
those already obtained in \cite{GarciaRecio:2010ki}. Again, we refer to that
work for further details and grounds on the identification proposed in Table
\ref{tab:1-12-0}. Very briefly, it looks quite natural to identify the first
two poles with the PDG $K^*_0(800)$ and $K^*_0(1430)$ states, in spite of
being the latter one much wider than the pole found in our scheme. The $K\pi$
branching fraction for this resonance is $93\%\pm 10\%$. For our pole at
$1425\MeV$, the direct coupling to $K\pi$ is not so dominant over the other
open channel, $\eta K$.  However, the couplings to the closed channels $K^*
\rho$, $K^* \omega$, $K^* \phi$ channels are much larger. As a consequence,
the resonance can decay into a virtual $K^*\rho$ pair and significantly
enhance the $K\pi$ decay probability, through the loop mechanism depicted in
the left panel of Fig.~\ref{fig:box}, thanks to the broad $\rho$ and $K^*$
spectral functions.

The identification of the third pole with the broad $K^*_0(1950)$ resonance is
less straightforward. Nevertheless, it should be pointed out that the
$K^*_0(1950)$ resonance is not firmly established yet and needs further
confirmation~\cite{Beringer:1900zz}.

A final comment is related with the employed UV cutoffs in this sector. Those
turn out to be of the order of $1\GeV$ in this case, as it was also the case
in the $(Y,I,J)=(0,0,0)$ sector, indicating that the dynamics of the
resonances in both sectors are mostly governed by the logs that appear in the
unitarity loops. This naturally explains why, for instance, the $K^*_0(800)$
is so wide, since it is placed well above the relevant threshold $K\pi$.
Indeed, this resonance is very similar to the $f_0(500)$, and it cannot be
interpreted as a Breit-Wigner narrow resonance.

\begin{table}[htpb]
\caption{Same as in Table \ref{tab:000}, but for the $(Y,I,J)=(1,1/2,0)$
  sector that corresponds to the $I(J^{P})=\frac{1}{2}(0^{+})$ quantum
  numbers. The subtraction constant has been set to $a=-1.52$ in all
  channels. The assignment of the third pole to the $K^*_0(1950)$ resonance is
  uncertain.}
\vspace{0.5cm}
\begin{tabular}{c|ccccc|c|cc}
\hline\hline
 $(M_R,\Gamma_R)$ & $K\pi$ & $\eta K$ & $K^*\rho$ & $K^*\omega$ &
$K^*\phi$ &
$(M_R,\Gamma_R)$~\cite{GarciaRecio:2010ki} & \multicolumn{2}{c}{PDG
  $(M_R,\Gamma_R)$~\cite{Beringer:1900zz}}  \\
 \hline
$(816,434)$ & {\bf 4.83} & 2.20 & 6.29 & 2.29 & 2.15 & $(812,347)$ &
 $K^*_0(800)$ & $(682\pm 29,547\pm24)$ \\
$(1425,54)$ & {\bf 1.91} & {\bf 1.02} & 8.11 & 10.69 & 5.70 &  $(1428,48)$ &
 $K^*_0(1430)$ & $(1425\pm50,270\pm80)$ \\
$(1782,90)$ & {\bf 0.06} & {\bf 2.92} & {\bf 0.68} & {\bf 1.07} & 12.21 & $(1787,74)$  &
 $K^*_0(1950)$ & $(1945\pm22,201\pm90)$ \\

\hline\hline
\end{tabular}\label{tab:1-12-0}
\end{table}
\subsection{Hypercharge 1, isospin 1/2 and spin 2}
\begin{table}[htpb]
\caption{Same as in Table \ref{tab:000}, but for 
the $(Y,I,J)=(1,1/2,2)$ sector that corresponds to the
$I(J^{P})=\frac{1}{2}(2^{+})$ quantum numbers. The
subtraction constant has been set to $a=-4.32$ in all channels. 
The assignment of the second pole  to
the $K^*_2(1980)$ resonance is  uncertain. } \label{tab:2-1/2-2} \vspace{0.5cm}
\begin{tabular}{c|ccc|c|cc}
\hline\hline
 $(M_R,\Gamma_R)$  & $ K^*\rho$ & $ K^*\omega$ & $ K^*\phi$ &
$(M_R,\Gamma_R)$~\cite{GarciaRecio:2010ki} & \multicolumn{2}{c}{PDG
  $(M_R,\Gamma_R)$~\cite{Beringer:1900zz}}  \\
 \hline
$(1430,0)$ & 6.30  & 4.23 & 6.81 & $(1701,313)$ & $K^*_2(1430)$&
 $(1429 \pm 4,104\pm 6)$  \\
$(1624,0)$ & 6.21 & 0.39 & 2.88 & & $K^*_2(1980)$&  $(1973 \pm 26,373\pm 70)$\\
\hline\hline
\end{tabular}
\end{table}
In this sector (Table~\ref{tab:2-1/2-2}), we find two poles in the FRS/SRS of
the amplitudes. In the PDG, two $K^*_2$ resonances [$K^*_2(1430)$ and
  $K^*_2(1980)$] are reported below $2\GeV$~\cite{Beringer:1900zz}, though
only the lightest one is firmly established. In the analysis of
Ref.~\cite{GarciaRecio:2010ki} just one state was found, and moreover, the
subtraction constants could not be fine-tuned to bring its mass below
$1.7\GeV$. The consideration of $\delta {\cal H}$ in the current approach
overcomes this problem, and it allows to generate a pole in the region of
$1430\MeV$. According to the PDG, the $K^*_2(1430)$ has a width of $104\pm
6\MeV$, in our approach we find a bound state, located below all the
thresholds. The PDG branching fractions are around 50\%, 25\%, 9\% and 3\% for
the $D$-wave modes $K\pi$, $K^*\pi$, $K\rho$ and $K\omega$, respectively. In
addition, the branching fraction of the $K^*\pi\pi$ channel is only about
13\%. Certainly, these modes can be also originated from the decay of the
resonance to $ K^*\rho$, $ K^*\omega$ and $ K^*\phi$ virtual pairs. In
particular, we expect the $K^*\rho$ channel to play an important role, because
of the broad spectral functions of both the $K^*$ and $\rho$ mesons and its
proximity to the mass of the resonance, since it can trigger a significant
part of the observed $K^*_2(1430)$ decays into $K\pi$ and $K^*\pi\pi$ (see
Fig.~\ref{fig:box}). Note that the large $K^*\phi$ coupling also provides a
contribution to the dominant $K\pi$ mode.

We should note, however, that we need to use values of the subtraction
constants that amount to UV cutoffs above $4\GeV$, which put some doubts on
the real nature of this state, as we discussed in $(Y,I,J)=$(0,0,2) and
(0,1,2) sectors. It might be the case that large UV cutoffs are needed to
compensate the genuine (non resonant driven) $D$-wave channels ignored in the
present coupled channels approach. In particular, in
Ref.~\cite{GarciaRecio:2010ki} it was already pointed out the possible
influence of the $D$-wave pseudoscalar-vector meson $K^* \pi$ channel, which
lies closer to the resonance mass than the pseudoscalar-pseudoscalar channels.

The hidden gauge approach of Ref.~\cite{Geng:2008gx} for $VV\to VV$ scattering
produces a resonance in this sector, with mass fine-tuned to $1430\MeV$, and
properties quite similar to those found here. There, all $D$-wave type
interactions were also ignored.

On the other hand, little is known about the $K_2^*(1980)$, and the assignment
of our second pole to this resonance is clearly uncertain. (Note that in
Ref.~\cite{GarciaRecio:2010ki}, a second pole was not generated in this
sector.) The large width of this state ($\Gamma \sim 400$) makes less
meaningful the difference between its mass and that of our pole, which might
be then associated to this resonance. Still, it should be noted again that the
$K_2^*(1980)$ resonance is not yet firmly established and needs further
confirmation. It might well happen that the pole obtained here corresponds to
a different state not yet detected.

\subsection{Exotics}

\begin{table}[htpb]
\caption{Same as in Table \ref{tab:000}, but for 
the $(Y,I,J)=(0,2,0)$ sector that corresponds to the exotic
$I^G(J^{PC})=2^+(0^{++})$ quantum numbers.  The subtraction constant has been set
to $a=-1.51$ in all channels. }\label{tab:X020} \vspace{0.5cm}
\begin{tabular}{c|cc|c|cc}
\hline\hline
 $(M_R,\Gamma_R)$& $\pi\pi$ & $\rho\rho$ &
$(M_R,\Gamma_R)$~\cite{GarciaRecio:2010ki} & \multicolumn{2}{c}{PDG
  $(M_R,\Gamma_R)$~\cite{Beringer:1900zz}} \\
 \hline
 $(1420,110)$ & {\bf 2.74} & 9.99 & $(1418,108)$ & $X(1420)$ & $(1420\pm20,160\pm10)$  \\
\hline\hline
\end{tabular}
\end{table}

Exotics refers here to meson states with quantum numbers that cannot be formed
by a $q\bar{q}$ pair. Quantum numbers with $I > 1$ or $|Y| > 1$ are
exotic. 

Besides the exotic poles with $J=1$ in the sectors $(Y=1, I=3/2)$ and $(Y=2,
I=0)$ already reported in Ref.~\cite{GarciaRecio:2010ki}, we find another
three exotic states with $J=0$ in region $1.4-1.6\GeV$. Their positions and
couplings are compiled in Tables \ref{tab:X020}, \ref{tab:1320} and
\ref{tab:210}. In these tables, we have also collected the pole positions
found in Ref.~\cite{GarciaRecio:2010ki}. These scalar exotic states appear in
the sectors $(Y=0, I=2)$, $(Y=1, I=3/2)$ and $(Y=2, I=1)$. As already pointed
out in Ref.~\cite{GarciaRecio:2010ki}, the matrices ${\mathcal D}_{\rm kin}$
and ${\mathcal D}_m$ in \Eq{vsu6} are identical in the three sectors. The
analogous statement holds for the new interactions ${\mathcal D}_2$ and
${\mathcal D}_3$ (\Eq{vsu6new}). Thus, clearly the three spin zero exotic
states are just related by flavor rotations.

\begin{table}[htpb]
\caption{Same as in Table \ref{tab:000}, but for 
  the $(Y,I,J)=(1,3/2,0)$ sector that corresponds to the exotic
 $I(J^{P})=\frac{3}{2}(0^{+})$ quantum numbers.  The subtraction constant has been set
to $a=-2.0$ in all channels. } \vspace{0.5cm}
\begin{tabular}{c|cc|c}
\hline\hline
 $(M_R,\Gamma_R)$ & $ K\pi$ & $ K^*\rho$ & $(M_R,\Gamma_R)$~\cite{GarciaRecio:2010ki} \\
 \hline
 $(1438,140)$ & {\bf 3.26} & 10.86 & $(1431,140)$\\
\hline\hline
\end{tabular}\label{tab:1320}
\end{table}

As in the $(Y,I,J)=(0,0,0)$ and $(Y,I,J)=(1,1/2,0)$ sectors, the inclusion of
$\delta{\cal H}$ in the present work has very little effect and the present
results are quite similar (practically identical) to those already obtained in
\cite{GarciaRecio:2010ki}. We refer to that work for further details as well
as for an overall picture of the poles with exotic quantum numbers predicted
for this SU(6) extension of the WT Lagrangian. In short, there is only state
listed in the PDG that can be associated with the exotic resonances predicted
by our model. This is the $X(1420)$ resonance, but it needs further
confirmation and its current evidence comes from a statistical
indication~\cite{Filippi:2000is} for a $\pi^+\pi^+$ resonant state in the
$\bar n p \to \pi^+ \pi^+ \pi^-$ annihilation reaction with data collected by
the OBELIX experiment. In our model, the pole is mainly a $\rho\rho$ bound
state with a small coupling to the $\pi\pi$ channel that moves the pole to the
SRS.  Within our scheme, the $\rho\rho\to \rho\rho$ amplitude is totally
symmetric under $I \leftrightarrow J$ exchange. As a consequence our
$\rho\rho$ potential in this sector ($I=2, J=0$) is the same as that in the
$I=0, J=2$ one. BSE amplitudes in both sectors will become different because
of coupled-channel and renormalization effects. Nevertheless, we expect the
$X(1420)$ to be the counterpart of the $f_2(1270)$, which appeared with a
large $\rho\rho$ spin two isoscalar component. As mentioned, the other two
spin zero exotic states in the $(Y=1, I=3/2)$ and $(Y=2, I=1)$ sectors should
be related to $X(1420)$ by a flavor rotation. However, there is no
experimental evidence of their existence yet.

\begin{table}[htpb]
\caption{Same as in Table \ref{tab:000}, but for 
the $(Y,I,J)=(2,1,0)$ sector that corresponds to the exotic
 $I(J^{P})=1(0^{+})$ quantum numbers.  The subtraction constant has been set
to $a=-2.0$ in all channels. } \vspace{0.5cm}
\begin{tabular}{c|cc|c}
\hline\hline
 $(M_R,\Gamma_R)$ & $\bar{K} K$ & $\bar{K^*} K^*$ & $(M_R,\Gamma_R)$~\cite{GarciaRecio:2010ki} \\
 \hline
$(1568,132)$ & {\bf 3.50} & 11.49 &$(1563,132)$ \\
\hline\hline
\end{tabular}\label{tab:210}
\end{table}

Finally, we just mention that the ${\mathcal D}_{\rm kin}$, ${\mathcal D}_m$,
${\mathcal D}_2$ and ${\mathcal D}_3$ matrices are identical in the three
sectors $(Y,I,J)=(0,2,2),(1,3/2,2)$ and $(2,1,2)$. They provide a repulsive
interaction and hence no resonance is predicted in those exotic sectors by our
model.

\section{Summary}
\label{sec:concl}

We have reviewed the model presented in Ref.~\cite{GarciaRecio:2010ki} to
address the dynamics of the low-lying even parity meson resonances. It is
based on a spin-flavor extension of the chiral WT Lagrangian, which is then
used to study the $S$-wave meson-meson interaction involving members not only
of the $\pi$-octet, but also of the $\rho$-nonet. Elastic unitarity in coupled
channels is restored by solving a renormalized coupled channels BSE, and a
certain pattern of SU(6) spin--flavor symmetry breaking is implemented. The
model probed to be phenomenologically successful in the $J^P=0^+$
and $1^+$ sectors. Actually in ~\cite{GarciaRecio:2010ki}, it was shown that
most of the low-lying even parity PDG meson resonances in these two spin
sectors could be classified according to multiplets of SU(6). However the
scheme of Ref.~\cite{GarciaRecio:2010ki} is not so successful for the
sectors with spin 2. It fails to appropriately describe some well established
$J^P=2^+$ resonances, like the $K^*_2(1430)$, that in the hidden gauge
formalism for vector mesons used in Ref.~\cite{Geng:2008gx} are dynamically
generated in a natural manner. In this work, we have improved on that by
supplementing the model of Ref.~\cite{GarciaRecio:2010ki} with new local $VV$
interactions consistent with CS.

To provide different pseudoscalar and vector mesons masses, a simple
spin-symmetry breaking local term that preserved CS was designed in
~\cite{GarciaRecio:2010ki}. Here, we have studied in detail the structure of
the SU(6) symmetry breaking local terms that respect (or softly break)
CS. Thus, in this work, we have derived the most general contact terms
consistent with the chiral symmetry breaking pattern of QCD as expressed in
terms of the field $U$. We have also shown that there is a finite number of
chirally invariant contact four meson-field interactions, restricted also by
the other symmetries of the problem. To reduce the number of parameters to a
manageable size, and in the spirit of large $N_C$, we have restricted our
analysis to interactions involving just one trace.

Further, we have carried out a phenomenological discussion of the effects of
these new terms. We find that their inclusion leads to a considerable
improvement of the description of the $J^P = 2^+$ sector, without spoiling the
main features of the predictions obtained in Ref.~\cite{GarciaRecio:2010ki}
for the $J^P = 0^+$ and $J^P = 1^+$ sectors. In particular, we have found a
significantly better description of the $I^G(J^{PC})=0^+(2^{++})$,
$1^-(2^{++})$ and the $I(J^{P})=\frac{1}{2}(2^{+})$ sectors, that correspond
to the $f_2(1270)$, $a_2(1320)$ and $K^*_2(1430)$ quantum numbers,
respectively. Besides the position of the resonances, we also estimate the
couplings of those resonances to the different channels, which is relevant to
describe the state structure and its favored decay modes. Our analysis shows
that $2^+$ states systematically require cutoff values which lie in the
boundary of their natural hadronic domain. This could be an indication that
$D$-wave mechanisms play some role in the formation of such
states. The fact that, in many cases, the thresholds of the main channels are
not too close to the resonance position, also suggests that pure $S$-wave
interactions would not necessarily saturate the formation mechanisms of those
resonances. Of course, for some particular mesonic resonances, it could also
be the case that they are mostly genuine rather than dynamically
generated. With this possible caveat in mind, we can say that the model
produces a rather robust and successful scheme to study the low-lying even
parity resonances that are dynamically generated by the  logs that appear
in the unitarity loops.

\begin{acknowledgments}
 This research was supported by the Spanish Ministerio de Econom\'\i a
 y Competitividad and European FEDER funds under the contracts 
   FIS2011-28853-C02-01, FIS2011-28853-C02-02, FIS2011-24149 and the Spanish
  Consolider-Ingenio 2010 Programme CPAN (CSD2007-00042),  by Generalitat
  Valenciana under contract PROMETEO/2009/0090, by Junta de
  Andaluc\'{\i}a grant FQM-225  and by the EU
  HadronPhysics2 project, grant agreement no. 227431. This work is
  also partly supported by the National Natural Science Foundation of
  China  under grant numbers 11005007 and 11105126. 
\end{acknowledgments}

\appendix
\section{Chiral invariant four meson interaction with a single trace}
\label{app:chiral}

In this appendix we  show that, the operators $\mathcal{O}_1$,
$\mathcal{O}_2$, $\mathcal{O}_3$ in \Eq{13}, already saturate the most general
chiral invariant interaction, modulo $O(\Phi_6^6)$, stemming from
single trace Lagrangian terms.

Rather than doing the expansion of the most general term
$\tr(\mathcal{U}_{ij}\mathcal{U}_{kl}\cdots)$ in powers of the meson field, we
just write down the possible operators in terms of the meson field and seek
the most general combination invariant under infinitesimal chiral rotations.
To alleviate the notation, we use $U_6=e^\phi$, $\phi$ being antihermitian and
dimensionless. This is related with the usual meson field by
$\phi=2i\Phi_6/f$.

The 8 possible terms, assuming other symmetries but not chiral
invariance, are as follows
\begin{eqnarray}
A_1 &=& \tr(\phi^2) 
,\nonumber \\
A_2 &=& \tr(\sigma_i \phi \sigma_i \phi) 
,\nonumber \\
A_3 &=& \tr(\phi^4) 
,\nonumber \\
A_4 &=& \tr(\sigma_i \phi \sigma_i \phi^3) 
,\nonumber \\
A_5 &=& \tr(\sigma_i \phi^2 \sigma_i \phi^2) 
,\nonumber \\
A_6 &=& i\epsilon_{ijk}\tr(\sigma_i \phi \sigma_j \phi \sigma_k \phi^2) 
,\nonumber \\
A_7 &=& \tr(\sigma_i \phi \sigma_i \phi \sigma_j \phi \sigma_j \phi) 
,\nonumber \\
A_8 &=& \tr(\sigma_i \phi \sigma_j \phi \sigma_i \phi \sigma_j \phi) 
.
\end{eqnarray}

The operators $A_1$ and $A_2$ give mass to the mesons, the other provide
interaction.

Under a chiral rotation $U_6 \to \Omega_L^\dagger U_6 \Omega_R$, and this
induces a non linear transformation on $\phi$. Vector invariance
($\Omega_L=\Omega_R$) is a similarity transformation which produces the same
transformation on $\phi$ and it is trivially satisfied by the 8
operators. Therefore we consider just axial transformations $\Omega_L=
\Omega_R^\dagger$. Only infinitesimal transformations are needed, $\Omega_R =
e^{\delta\alpha/2} = 1+\frac{1}{2}\delta\alpha$, with $\delta\alpha$
infinitesimal, antihermitian and spinless. This induces the transformation
\begin{equation}
\delta\phi = \delta\alpha 
+ \frac{1}{12}\delta\alpha \phi^2 
+ \frac{1}{12}\phi^2 \delta \alpha 
- \frac{1}{6} \phi \delta\alpha \phi
+ O(\phi^4)
.  
\end{equation}
(To all orders in the meson field, the infinitesimal axial variation contains
only even powers of $\phi$.)

The variations of $A_1$ and $A_2$ produce terms of $O(\phi)$ that can
only be canceled by choosing a suitable combination of the two
operators. Also they produce terms of $O(\phi^3)$. They should cancel
with the corresponding variations from the quartic terms, taking
suitable combinations of them. The cancellation to order $O(\phi^5)$
is of no concern to us as this involves $O(\phi^6)$ interactions.  The
cancellation will be automatic for the expansion of any of the terms
$\tr(\mathcal{U}_{ij}\mathcal{U}_{kl}\cdots)$ since chiral invariance
is manifest in those terms.

For a generic operator $H=\sum_{i=1}^8 c_i A_i$, the condition $\delta H =
O(\phi^5)$ gives the following conditions
\begin{eqnarray}
0 &=& 2c_1 + 6 c_2 
,\nonumber \\
0 &=& 4c_3 + 3 c_4 
,\nonumber \\
0 &=& -\frac{1}{3} c_2 + c_4 + 8 c_7 + 4 c_8 
,\nonumber \\
0 &=& \frac{1}{6} c_2 + c_4 + 2 c_5 - 2 c_6 
,\nonumber \\
0 &=& \frac{1}{6} c_2 + c_4 + 2 c_5 - 2 c_6 
,\nonumber \\
0 &=& 2 c_6 - 2 c_7 + 4 c_8
.
\end{eqnarray}
They correspond, respectively, to the vanishing of the coefficients of
$\tr(\phi\delta\alpha)$, $\tr(\phi^3\delta\alpha)$,
$\tr(\phi\sigma_i\phi\sigma_i\phi\delta\alpha)$,
$\tr(\sigma_i\phi\sigma_i\phi^2\delta\alpha)$,
$\tr(\sigma_i\phi^2\sigma_i\phi\delta\alpha)$, and
$i\epsilon_{ijk}\tr(\sigma_i\phi\sigma_j\phi\sigma_k\phi\delta\alpha)$.

The 5 independent relations leave 3 chiral invariant combinations. They can be
taken as
\begin{equation}
H_{\mathrm{inv}} = 
c_1(A_1-\frac{1}{3}A_2-\frac{1}{36}A_6-\frac{1}{60}A_7+\frac{1}{180}A_8)
+
c_3(A_3-\frac{4}{3}A_4-\frac{2}{3}A_6+\frac{1}{3}A_8)
+
c_5(A_5 + A_6 +\frac{1}{5}A_7-\frac{2}{5}A_8)
.
\end{equation}
The three combinations $\mathcal{O}_1$, $(\mathcal{O}_2-20\mathcal{O}_1)/12$,
and $\mathcal{O}_3/3$ in Section \ref{sec:su6new} correspond, respectively, to
$(c_1,c_3,c_5) = (3,\frac{1}{4},\frac{1}{4})$, $(0,1,\frac{5}{6})$, and
$(0,1,0)$.

\section{Coefficients of the $S$-wave tree level amplitudes}
\label{app:tables}

This Appendix gives the ${\mathcal D}_{2}$ and ${\mathcal D}_{3}$ matrices of
the $S$-wave tree level meson-meson amplitudes in \Eq{vsu6new}, for the
various $YIJ$ sectors (Tables~\ref{tab:initial}-\ref{tab:final}). Note that
for the $Y=0$ channels, $G$-parity is conserved, and that all $Y=0$ states
have well-defined $G$-parity except the $\bar K^* K $ and $ K^* \bar K $
states, but the combinations $\left(\bar K K^* \pm K \bar K^* \right)/\sqrt 2$
are actually $G$-parity eigenstates with eigenvalues $\pm 1$. These states
will be denoted $(\bar{K}K^*)_S$ and $(\bar{K}K^*)_A$, respectively.
\newpage
\subsection{${\mathcal D}_{2}$}
\begin{table}[htpb]
      \renewcommand{\arraystretch}{2}
     \setlength{\tabcolsep}{0.4cm}
     \caption{$(Y,I,J)=(0,0,0)$.}
\begin{tabular}{c|cccccccc}
\hline\hline
 & $\pi\pi$ &  $\bar{K}K$ & $\eta\eta$ & $\rho\rho$ & $\omega\omega$ & $\omega\phi$  & $\bar{K}^* K^*$ &  $\phi\phi$ \\
\hline 
 & 0 & 0 & 0 & 0 & 0 & 0 & 0 & 0 \\
 & 0 & 0 & 0 & 0 & 0 & 0 & 0 & 0 \\
 & 0 & 0 & 0 & 0 & 0 & 0 & 0 & 0 \\
 & 0 & 0 & 0 & $-\frac{208}{3}$ & $\frac{80}{\sqrt{3}}$ & 0 & $24 \sqrt{3}$ & 0 \\
 & 0 & 0 & 0 & $\frac{80}{\sqrt{3}}$ & $-\frac{80}{3}$  & 0 & $-24$ & 0 \\
 & 0 & 0 & 0 & 0 & 0 & 0 & $\frac{16}{3}$ & 0 \\
 & 0 & 0 & 0 &$24 \sqrt{3}$ & $-24$ & $ \frac{16}{3}$ & $-72$ & $-48$ \\
 & 0 & 0 & 0 & 0 & 0 & 0 & $-48$ & $-\frac{160}{3}$\\
\hline\hline
\end{tabular}
\label{tab:initial}
\end{table}

\begin{table}[htpb]
      \renewcommand{\arraystretch}{2}
     \setlength{\tabcolsep}{0.4cm}
     \caption{$(Y,I,J)=(0,0,1)$.}
\begin{tabular}{c|cccccccc}
\hline\hline$G$ & $\eta\phi$ &  $\eta\omega$ & $\pi\rho$ & $(\bar{K}K^*)_A$ & $\bar{K}^* K^*$ & $\omega\phi$ &    $(\bar{K}K^*)_S$   \\\hline 
$-$ & 0 & 0 & 0 & 0 & 0 &  &  \\
$-$ & 0 & 0 & 0 & 0 & 0 &  &  \\
$-$ & 0 & 0 & 0 & 0 & 0 &  &  \\
$-$ & 0 & 0 & 0 & 0 & 0 &  &  \\
$-$ & 0 & 0 & 0 & 0 & $-28$ &  &  \\
$+$ &  &  &  &  &  & 0 & 0 \\
$+$ &  &  &  &  &  & 0 & 0 \\
\hline\hline
\end{tabular}
\end{table}

\begin{table}[htpb]
      \renewcommand{\arraystretch}{2}
     \setlength{\tabcolsep}{0.4cm}
     \caption{$(Y,I,J)=(0,0,2)$.}
\begin{tabular}{c|ccccc}
\hline\hline
 & $\rho\rho$ & $\omega\omega$& $\omega\phi$ &  $\bar{K}^* K^*$ &  $\phi\phi$ \\\hline 
 & $-\frac{16}{3}$ & $\frac{32}{\sqrt{3}}$ & 0 & $4 \sqrt{3}$ & 0 \\
 & $\frac{32}{\sqrt{3}}$ & $-\frac{32}{3}$ & 0 & $-4$ & 0 \\
 & 0 & 0 & 0 & $\frac{40}{3}$ & 0 \\
 & $4 \sqrt{3}$ & $-4$ & $\frac{40}{3}$ & $-12$ & $-8$ \\
 & 0 & 0 & 0 & $-8$ & $-\frac{64}{3}$ \\
\hline\hline
\end{tabular}
\end{table}

\begin{table}[htpb]
      \renewcommand{\arraystretch}{2}
     \setlength{\tabcolsep}{0.4cm}
     \caption{$(Y,I,J)=(0,1,0)$.}
\begin{tabular}{c|ccccc}
\hline\hline
 & $\eta\pi$ & $\bar{K}K$ & $\omega\rho$ & $\phi\rho$& $\bar{K}^*K^*$  \\\hline 
 & 0 & 0 & 0 & 0 & 0 \\
 & 0 & 0 & 0 & 0 & 0 \\
 & 0 & 0 & $-\frac{160}{3}$ & 0 & $24\sqrt{2}$ \\
 & 0 & 0 & 0 & 0 & $-\frac{16}{3}$ \\
 & 0 & 0 & $24\sqrt{2}$ & $-\frac{16}{3}$ & $-24$ \\
\hline\hline
\end{tabular}
\end{table}

\begin{table}[htpb]
      \renewcommand{\arraystretch}{2}
     \setlength{\tabcolsep}{0.2cm}
     \caption{$(Y,I,J)=(0,1,1)$.}
\begin{tabular}{c|cccccccccc}
\hline\hline
 $G$ & $\pi \phi$ & $\pi \omega$ & $\eta\rho$ 
& $(\bar{K}K^*)_S$ & $\rho\rho$  &  $\bar{K}^* K^*$ 
& $\pi\rho$ & $(\bar{K}K^*)_A$ & $\omega\rho$ & $\phi\rho$ \\
\hline 
 $+$& 0 & 0 & 0 & 0 & 0 & 0 &  &  &  &  \\
 $+$& 0 & 0 & 0 & 0 & 0 & 0 &  &  &  &  \\
 $+$& 0 & 0 & 0 & 0 & 0 & 0 &  &  &  &  \\
 $+$& 0 & 0 & 0 & 0 & 0 & 0 &  &  &  &  \\
 $+$& 0 & 0 & 0 & 0 & $-\frac{56}{3}$ & $\frac{28 \sqrt{2}}{3}$ &  &  &  &  \\
 $+$& 0 & 0 & 0 & 0 & $\frac{28 \sqrt{2}}{3}$ & $-\frac{28}{3}$ &  &  &  &  \\
 $-$&  &  &  &  &  &  & 0 & 0 & 0 & 0 \\
 $-$&  &  &  &  &  &  & 0 & 0 & 0 & 0 \\
 $-$&  &  &  &  &  &  & 0 & 0 & 0 & 0 \\
 $-$&  &  &  &  &  &  & 0 & 0 & 0 & 0 \\
\hline\hline
\end{tabular}
\end{table}

\begin{table}[htpb]
      \renewcommand{\arraystretch}{2}
     \setlength{\tabcolsep}{0.2cm}
     \caption{$(Y,I,J)=(0,1,2)$.}
\begin{tabular}{c|ccc}
\hline\hline
 & $\omega\rho$ & $\phi\rho$  & $\bar{K}^* K^*$\\\hline 
 & $-\frac{64}{3}$  & 0 & $4 \sqrt{2}$ \\
 & 0 & 0 & $-\frac{40}{3}$ \\
 & $4 \sqrt{2}$ & $-\frac{40}{3}$ & $-4$ \\
\hline\hline
\end{tabular}
\end{table}

\begin{table}[htpb]
      \renewcommand{\arraystretch}{2}
     \setlength{\tabcolsep}{0.2cm}
     \caption{$(Y,I,J)=(0,2,0)$.} \label{tab:020m}
\begin{tabular}{c|cc}
\hline\hline
 & $\pi\pi$ & $\rho\rho$ \\\hline 
 & 0 & 0 \\
 & 0 & $-\frac{16}{3}$\\
\hline\hline
\end{tabular}
\end{table}

\begin{table}[htpb]
      \renewcommand{\arraystretch}{2}
     \setlength{\tabcolsep}{0.2cm}
     \caption{$(Y,I,J)=(0,2,1)$.}
\begin{tabular}{c|c}
\hline\hline
 & $\pi\rho$  \\\hline 
 & 0\\
\hline\hline
\end{tabular}
\end{table}

\begin{table}[htpb]
      \renewcommand{\arraystretch}{2}
     \setlength{\tabcolsep}{0.2cm}
     \caption{$(Y,I,J)=(0,2,2)$.}
\begin{tabular}{c|c}
\hline\hline
 & $\rho\rho$  \\\hline 
 & $-\frac{40}{3}$\\
\hline\hline
\end{tabular}
\end{table}

\begin{table}[htpb]
      \renewcommand{\arraystretch}{2}
     \setlength{\tabcolsep}{0.2cm}
     \caption{$(Y,I,J)=(1,1/2,0)$.}
\begin{tabular}{c|ccccc}
\hline\hline
& $K\pi$ & $\eta K$ & $K^*\rho$ & $K^*\omega$ & $K^*\phi$\\\hline
& 0 & 0 & 0 & 0 & 0 \\
& 0 & 0 & 0 & 0 & 0 \\
& 0 & 0 & $-\frac{100}{3}$ & $-\frac{44}{\sqrt{3}}$ & $12 \sqrt{6}$ \\
& 0 & 0 & $-\frac{44}{\sqrt{3}}$ & $-\frac{44}{3}$ & $12 \sqrt{2}$ \\
& 0 & 0 & $12 \sqrt{6}$ & $12 \sqrt{2}$ & $-\frac{88}{3}$\\
\hline\hline
\end{tabular}
\end{table}

\begin{table}[htpb]
      \renewcommand{\arraystretch}{2}
     \setlength{\tabcolsep}{0.2cm}
     \caption{$(Y,I,J)=(1,1/2,1)$.}
\begin{tabular}{c|cccccccc}
\hline\hline
 & $\pi K^*$ &  $ K \rho$ & $ K \omega$ & $\eta K^*$ & $ K\phi$  & $ K^*\rho$ & $ K^*\omega$ & $ K^*\phi$  \\\hline 
 & 0 & 0 & 0 & 0 & 0 & 0 & 0 & 0 \\
 & 0 & 0 & 0 & 0 & 0 & 0 & 0 & 0 \\
 & 0 & 0 & 0 & 0 & 0 & 0 & 0 & 0 \\
 & 0 & 0 & 0 & 0 & 0 & 0 & 0 & 0 \\
 & 0 & 0 & 0 & 0 & 0 & 0 & 0 & 0 \\
 & 0 & 0 & 0 & 0 & 0 & $-14$ & $-\frac{14}{\sqrt{3}}$ & $-14 \sqrt{\frac{2}{3}}$ \\
 & 0 & 0 & 0 & 0 & 0 & $-\frac{14}{\sqrt{3}}$ & $-\frac{14}{3}$ & $-\frac{14 \sqrt{2}}{3}$ \\
 & 0 & 0 & 0 & 0 & 0 & $-14 \sqrt{\frac{2}{3}}$ & $-\frac{14 \sqrt{2}}{3}$ & $-\frac{28}{3}$ \\
\hline\hline
\end{tabular}
\end{table}

\begin{table}[htpb]
      \renewcommand{\arraystretch}{2}
     \setlength{\tabcolsep}{0.2cm}
     \caption{$(Y,I,J)=(1,1/2,2)$.}
\begin{tabular}{c|ccc}
\hline\hline
 & $ K^*\rho$ & $ K^*\omega$ & $ K^*\phi$  \\\hline 
 & $\frac{2}{3}$ & $-\frac{26}{\sqrt{3}}$ & $2 \sqrt{6}$ \\
 & $-\frac{26}{\sqrt{3}}$ & $-\frac{26}{3}$ & $2 \sqrt{2}$ \\
 & $2 \sqrt{6}$ & $2 \sqrt{2}$ & $-\frac{52}{3}$\\
\hline\hline
\end{tabular}
\end{table}

\begin{table}[htpb]
      \renewcommand{\arraystretch}{2}
     \setlength{\tabcolsep}{0.2cm}
     \caption{$(Y,I,J)=(1,3/2,0)$.}
\begin{tabular}{c|ccc}
\hline\hline
 & $ K\pi$ & $ K^*\rho$  \\\hline 
 & $0$ & 0 \\
 &  0  & $-\frac{16}{3}$\\
\hline\hline
\end{tabular}
\end{table}

\begin{table}[htpb]
      \renewcommand{\arraystretch}{2}
     \setlength{\tabcolsep}{0.2cm}
     \caption{$(Y,I,J)=(1,3/2,1)$.}
\begin{tabular}{c|ccc}
\hline\hline
& $\pi K^*$ & $ K\rho$ & $ K^*\rho$  \\\hline 
& 0 & 0 & 0 \\
& 0 & 0 & 0 \\
& 0 & 0 & 0 \\
\hline\hline
\end{tabular}
\end{table}

\begin{table}[htpb]
      \renewcommand{\arraystretch}{2}
     \setlength{\tabcolsep}{0.2cm}
     \caption{$(Y,I,J)=(1,3/2,2)$.}
\begin{tabular}{c|c}
\hline\hline
 & $K^*\rho $  \\\hline
 & $-\frac{40}{3}$\\ 
\hline\hline
\end{tabular}
\end{table}

\clearpage

\begin{table}[htpb]
      \renewcommand{\arraystretch}{2}
     \setlength{\tabcolsep}{0.2cm}
     \caption{$(Y,I,J)=(2,0,1)$.}
\begin{tabular}{c|cc}
\hline\hline
 & $ K K^*$ & $K^* K^*$  \\\hline
 & 0 & 0 \\
 & 0 & 0\\
\hline\hline
\end{tabular}
\end{table}

\begin{table}[htpb]
      \renewcommand{\arraystretch}{2}
     \setlength{\tabcolsep}{0.2cm}
     \caption{$(Y,I,J)=(2,1,0)$.}
\begin{tabular}{c|cc}
\hline\hline
 & $K K$ & $K^* K^*$  \\\hline
 & $0$ & $0$ \\
 & $0$ & $-\frac{16}{3}$\\
\hline\hline
\end{tabular}
\end{table}

\begin{table}[htpb]
      \renewcommand{\arraystretch}{2}
     \setlength{\tabcolsep}{0.2cm}
     \caption{$(Y,I,J)=(2,1,1)$.}
\begin{tabular}{c|cc}
\hline\hline
 & $ K K^*$ \\\hline
 & $0$  \\
\hline\hline
\end{tabular}
\end{table}

\begin{table}[htpb]
      \renewcommand{\arraystretch}{2}
     \setlength{\tabcolsep}{0.2cm}
     \caption{$(Y,I,J)=(2,1,2)$.}
\begin{tabular}{c|c}
\hline\hline
 &  $K^* K^*$  \\\hline
 &  $-\frac{40}{3}$  \\
\hline\hline
\end{tabular}
\end{table}

\clearpage

\subsection{${\mathcal D}_{3}$}
\begin{table}[htpb]
      \renewcommand{\arraystretch}{2}
     \setlength{\tabcolsep}{0.4cm}
     \caption{$(Y,I,J)=(0,0,0)$.}
\begin{tabular}{c|cccccccc}
\hline\hline
 & $\pi\pi$ &  $\bar{K}K$ & $\eta\eta$ & $\rho\rho$ & $\omega\omega$ & $\omega\phi$  & $\bar{K}^* K^*$ &  $\phi\phi$ \\
\hline 
 & 0 & 0 & 0 & 0 & 0 & 0 & 0 & 0 \\
 & 0 & 0 & 0 & 0 & 0 & 0 & 0 & 0 \\
 & 0 & 0 & 0 & 0 & 0 & 0 & 0 & 0 \\
 & 0 & 0 & 0 & $-\frac{128}{3}$ & 0 & 0 & $\frac{32}{\sqrt{3}}$ & 0 \\
 & 0 & 0 & 0 & 0 & 0  & 0 & $-\frac{32}{3}$ & 0 \\
 & 0 & 0 & 0 & 0 & 0 & 0 & $-\frac{64}{3}$ & 0 \\
 & 0 & 0 & 0 &$\frac{32}{\sqrt{3}} $ & $-\frac{32}{3}$ & $ -\frac{64}{3}$ & $-32$ & $-\frac{64}{3}$ \\
 & 0 & 0 & 0 & 0 & 0 & 0 & $-\frac{64}{3}$ & 0\\
\hline\hline
\end{tabular}
\end{table}

\begin{table}[htpb]
      \renewcommand{\arraystretch}{2}
     \setlength{\tabcolsep}{0.4cm}
     \caption{$(Y,I,J)=(0,0,1)$.}
\begin{tabular}{c|cccccccc}
\hline\hline$G$ & $\eta\phi$ &  $\eta\omega$ & $\pi\rho$ & $(\bar{K}K^*)_A$ & $\bar{K}^* K^*$ & $\omega\phi$ &    $(\bar{K}K^*)_S$   \\\hline 
$-$ & 0 & 0 & 0 & 0 & 0 &  &  \\
$-$ & 0 & 0 & 0 & 0 & 0 &  &  \\
$-$ & 0 & 0 & 0 & 0 & 0 &  &  \\
$-$ & 0 & 0 & 0 & 0 & 0 &  &  \\
$-$ & 0 & 0 & 0 & 0 & $-48$ &  &  \\
$+$ &  &  &  &  &  & 0 & 0 \\
$+$ &  &  &  &  &  & 0 & 0 \\
\hline\hline
\end{tabular}
\end{table}

\begin{table}[htpb]
      \renewcommand{\arraystretch}{2}
     \setlength{\tabcolsep}{0.4cm}
     \caption{$(Y,I,J)=(0,0,2)$.}
\begin{tabular}{c|ccccc}
\hline\hline
 & $\rho\rho$ & $\omega\omega$& $\omega\phi$ &  $\bar{K}^* K^*$ &  $\phi\phi$ \\\hline 
 & $\frac{64}{3}$ & 0 & 0 & $-\frac{16}{\sqrt{3}}$ & 0 \\
 & 0 & 0 & 0 & $\frac{16}{3}$ & 0 \\
 & 0 & 0 & 0 & $\frac{32}{3}$ & 0 \\
 & $-\frac{16}{\sqrt{3}}$ & $\frac{16}{3}$ & $\frac{32}{3}$ & $16$ & $\frac{32}{3}$ \\
 & 0 & 0 & 0 & $\frac{32}{3}$ & 0 \\
\hline\hline
\end{tabular}
\end{table}

\begin{table}[htpb]
      \renewcommand{\arraystretch}{2}
     \setlength{\tabcolsep}{0.4cm}
     \caption{$(Y,I,J)=(0,1,0)$.}
\begin{tabular}{c|ccccc}
\hline\hline
 & $\eta\pi$ & $\bar{K}K$ & $\omega\rho$ & $\phi\rho$& $\bar{K}^*K^*$  \\\hline 
 & 0 & 0 & 0 & 0 & 0 \\
 & 0 & 0 & 0 & 0 & 0 \\
 & 0 & 0 & 0 & 0 & $\frac{32\sqrt{2}}{3}$ \\
 & 0 & 0 & 0 & 0 & $\frac{64}{3}$ \\
 & 0 & 0 & $\frac{32\sqrt{2}}{3}$ & $\frac{64}{3}$ & $-\frac{32}{3}$ \\
\hline\hline
\end{tabular}
\end{table}

\begin{table}[htpb]
      \renewcommand{\arraystretch}{2}
     \setlength{\tabcolsep}{0.2cm}
     \caption{$(Y,I,J)=(0,1,1)$.}
\begin{tabular}{c|cccccccccc}
\hline\hline
 $G$ & $\pi \phi$ & $\pi \omega$ & $\eta\rho$ 
& $(\bar{K}K^*)_S$ & $\rho\rho$  &  $\bar{K}^* K^*$ 
& $\pi\rho$ & $(\bar{K}K^*)_A$ & $\omega\rho$ & $\phi\rho$ \\
\hline 
 $+$& 0 & 0 & 0 & 0 & 0 & 0 &  &  &  &  \\
 $+$& 0 & 0 & 0 & 0 & 0 & 0 &  &  &  &  \\
 $+$& 0 & 0 & 0 & 0 & 0 & 0 &  &  &  &  \\
 $+$& 0 & 0 & 0 & 0 & 0 & 0 &  &  &  &  \\
 $+$& 0 & 0 & 0 & 0 & $-32$ & $16 \sqrt{2}$ &  &  &  &  \\
 $+$& 0 & 0 & 0 & 0 & $16 \sqrt{2}$ & $-16$ &  &  &  &  \\
 $-$&  &  &  &  &  &  & 0 & 0 & 0 & 0 \\
 $-$&  &  &  &  &  &  & 0 & 0 & 0 & 0 \\
 $-$&  &  &  &  &  &  & 0 & 0 & 0 & 0 \\
 $-$&  &  &  &  &  &  & 0 & 0 & 0 & 0 \\
\hline\hline
\end{tabular}
\end{table}

\begin{table}[htpb]
      \renewcommand{\arraystretch}{2}
     \setlength{\tabcolsep}{0.2cm}
     \caption{$(Y,I,J)=(0,1,2)$.}
\begin{tabular}{c|ccc}
\hline\hline
 & $\omega\rho$ & $\phi\rho$  & $\bar{K}^* K^*$\\\hline 
 & 0  & 0 & $-\frac{16 \sqrt{2}}{3}$ \\
 & 0 & 0 & $-\frac{32}{3}$ \\
 & $-\frac{16 \sqrt{2}}{3}$ & $-\frac{32}{3}$ & $\frac{16}{3}$ \\
\hline\hline
\end{tabular}
\end{table}

\begin{table}[htpb]
      \renewcommand{\arraystretch}{2}
     \setlength{\tabcolsep}{0.2cm}
     \caption{$(Y,I,J)=(0,2,0)$.} 
\begin{tabular}{c|cc}
\hline\hline
 & $\pi\pi$ & $\rho\rho$ \\\hline 
 & 0 & 0 \\
 & 0 & $\frac{64}{3}$\\
\hline\hline
\end{tabular}
\end{table}

\begin{table}[htpb]
      \renewcommand{\arraystretch}{2}
     \setlength{\tabcolsep}{0.2cm}
     \caption{$(Y,I,J)=(0,2,1)$.}
\begin{tabular}{c|c}
\hline\hline
 & $\pi\rho$  \\\hline 
 & 0\\
\hline\hline
\end{tabular}
\end{table}

\begin{table}[htpb]
      \renewcommand{\arraystretch}{2}
     \setlength{\tabcolsep}{0.2cm}
     \caption{$(Y,I,J)=(0,2,2)$.}
\begin{tabular}{c|c}
\hline\hline
 & $\rho\rho$  \\\hline 
 & $-\frac{32}{3}$\\
\hline\hline
\end{tabular}
\end{table}

\begin{table}[htpb]
      \renewcommand{\arraystretch}{2}
     \setlength{\tabcolsep}{0.2cm}
     \caption{$(Y,I,J)=(1,1/2,0)$.}
\begin{tabular}{c|ccccc}
\hline\hline
& $K\pi$ & $\eta K$ & $K^*\rho$ & $K^*\omega$ & $K^*\phi$\\\hline
& 0 & 0 & 0 & 0 & 0 \\
& 0 & 0 & 0 & 0 & 0 \\
& 0 & 0 & $-\frac{80}{3}$ & $\frac{16}{\sqrt{3}}$ & $16 \sqrt{\frac23}$ \\
& 0 & 0 & $\frac{16}{\sqrt{3}}$ & $\frac{16}{3}$ & $\frac{16 \sqrt{2}}{3}$ \\
& 0 & 0 & $16 \sqrt{\frac23}$ & $\frac{16 \sqrt{2}}{3}$ & $\frac{32}{3}$\\
\hline\hline
\end{tabular}
\end{table}
\begin{table}[htpb]
      \renewcommand{\arraystretch}{2}
     \setlength{\tabcolsep}{0.2cm}
     \caption{$(Y,I,J)=(1,1/2,1)$.}
\begin{tabular}{c|cccccccc}
\hline\hline
 & $\pi K^*$ &  $ K \rho$ & $ K \omega$ & $\eta K^*$ & $ K\phi$  & $ K^*\rho$ & $ K^*\omega$ & $ K^*\phi$  \\\hline 
 & 0 & 0 & 0 & 0 & 0 & 0 & 0 & 0 \\
 & 0 & 0 & 0 & 0 & 0 & 0 & 0 & 0 \\
 & 0 & 0 & 0 & 0 & 0 & 0 & 0 & 0 \\
 & 0 & 0 & 0 & 0 & 0 & 0 & 0 & 0 \\
 & 0 & 0 & 0 & 0 & 0 & 0 & 0 & 0 \\
 & 0 & 0 & 0 & 0 & 0 & $-24$ & $-8\sqrt{3}$ & $-8\sqrt{6}$ \\
 & 0 & 0 & 0 & 0 & 0 & $-8\sqrt{3}$ & $-8$ & $-8\sqrt{2}$ \\
 & 0 & 0 & 0 & 0 & 0 & $-8\sqrt{6}$ & $-8\sqrt{2}$ & $-16$ \\
\hline\hline
\end{tabular}
\end{table}

\begin{table}[htpb]
      \renewcommand{\arraystretch}{2}
     \setlength{\tabcolsep}{0.2cm}
     \caption{$(Y,I,J)=(1,1/2,2)$.}
\begin{tabular}{c|ccc}
\hline\hline
 & $ K^*\rho$ & $ K^*\omega$ & $ K^*\phi$  \\\hline 
 & $\frac{40}{3}$ & $-\frac{8}{\sqrt{3}}$ & $-8 \sqrt{\frac23}$ \\
 & $-\frac{8}{\sqrt{3}}$ & $-\frac{8}{3}$ & $-\frac{8\sqrt{2}}{3}$ \\
 & $-8 \sqrt{\frac23}$ & $-\frac{8\sqrt{2}}{3}$ & $-\frac{16}{3}$\\
\hline\hline
\end{tabular}
\end{table}

\begin{table}[htpb]
      \renewcommand{\arraystretch}{2}
     \setlength{\tabcolsep}{0.2cm}
     \caption{$(Y,I,J)=(1,3/2,0)$.}
\begin{tabular}{c|ccc}
\hline\hline
 & $ K\pi$ & $ K^*\rho$  \\\hline 
 & $0$ & 0 \\
 &  0  & $\frac{64}{3}$\\
\hline\hline
\end{tabular}
\end{table}

\begin{table}[htpb]
      \renewcommand{\arraystretch}{2}
     \setlength{\tabcolsep}{0.2cm}
     \caption{$(Y,I,J)=(1,3/2,1)$.}
\begin{tabular}{c|ccc}
\hline\hline
& $\pi K^*$ & $ K\rho$ & $ K^*\rho$  \\\hline 
& 0 & 0 & 0 \\
& 0 & 0 & 0 \\
& 0 & 0 & 0 \\
\hline\hline
\end{tabular}
\end{table}

\begin{table}[htpb]
      \renewcommand{\arraystretch}{2}
     \setlength{\tabcolsep}{0.2cm}
     \caption{$(Y,I,J)=(1,3/2,2)$.}
\begin{tabular}{c|c}
\hline\hline
 & $K^*\rho $  \\\hline
 & $-\frac{32}{3}$\\ 
\hline\hline
\end{tabular}
\end{table}

\clearpage
\begin{table}[htpb]
      \renewcommand{\arraystretch}{2}
     \setlength{\tabcolsep}{0.2cm}
     \caption{$(Y,I,J)=(2,0,1)$.}
\begin{tabular}{c|cc}
\hline\hline
 & $ K K^*$ & $K^* K^*$  \\\hline
 & 0 & 0 \\
 & 0 & 0\\
\hline\hline
\end{tabular}
\end{table}

\begin{table}[htpb]
      \renewcommand{\arraystretch}{2}
     \setlength{\tabcolsep}{0.2cm}
     \caption{$(Y,I,J)=(2,1,0)$.}
\begin{tabular}{c|cc}
\hline\hline
 & $K K$ & $K^* K^*$  \\\hline
 & $0$ & $0$ \\
 & $0$ & $\frac{64}{3}$\\
\hline\hline
\end{tabular}
\end{table}

\begin{table}[htpb]
      \renewcommand{\arraystretch}{2}
     \setlength{\tabcolsep}{0.2cm}
     \caption{$(Y,I,J)=(2,1,1)$.}
\begin{tabular}{c|cc}
\hline\hline
 & $ K K^*$ \\\hline
 & $0$  \\
\hline\hline
\end{tabular}
\end{table}

\begin{table}[htpb]
      \renewcommand{\arraystretch}{2}
     \setlength{\tabcolsep}{0.2cm}
     \caption{$(Y,I,J)=(2,1,2)$.}
\begin{tabular}{c|c}
\hline\hline
 &  $K^* K^*$  \\\hline
 &  $-\frac{32}{3}$  \\
\hline\hline
\end{tabular}
\label{tab:final}
\end{table}

%

\end{document}